\documentclass[journal,twocolumn]{IEEEtran}

\IEEEoverridecommandlockouts
\usepackage{color}
\usepackage{subfigure,bbm}
\usepackage{psfrag}
\usepackage{cite}
\usepackage{graphicx,xcolor}
\usepackage{amsmath}
\usepackage{hyperref}
\usepackage{multirow,multicol,relsize}
\usepackage{array,hhline}
\usepackage{epsfig}
\usepackage{epstopdf}
\usepackage{balance}
\usepackage{enumerate}
\usepackage{graphicx,cite}
\usepackage{times}
\usepackage{amssymb,lipsum}
\usepackage{algorithm}
\usepackage{algorithmic,booktabs}
\usepackage{makecell}

\newtheorem{Lemma}{Lemma}
\newtheorem{Theorem}{Theorem}

\newtheorem{Remark}{Remark}

\newtheorem{Corollary}{Corollary}
\newtheorem{Proposition}{Proposition}

\begin{document}

\title{Exploiting In-Slot\\Micro-Synchronism for S-ALOHA}

\author{Yangqian Hu, Jun-Bae Seo, {\em  Member, IEEE}, and Hu Jin, {\em  Senior Member}, {\em IEEE}% <-this % stops a space
\thanks{ Y. Hu and H. Jin are with the Division of Electrical and Electronic Engineering, Hanyang University, Ansan 15588, South Korea (e-mail: yijie86@hanyang.ac.kr, hjin@hanyang.ac.kr).}
\thanks{J.-B. Seo is with the Department of Information and Communication Engineering, Gyeongsang National University,  Tongyeong 53064, South Korea, (e-mail: jbseo@gnu.ac.kr).}% <-this % stops an unwanted space
}

%\markboth{Journal of \LaTeX\ Class Files,~Vol.~14, No.~8, August~2015}
%{Shell \MakeLowercase{\textit{et al.}}: Bare Demo of IEEEtran.cls for Computer Society Journals}

\maketitle 
 \begin{abstract}
%Proliferation of the urban Internet-of-Things (IoTs) for smart cities has fuelled massive amounts of data over wireless cellular networks. Upcoming  random access (RA) system of wireless cellular networks, e.g., 5G New Radio (NR), based on Slotted ALOHA (S-ALOHA) system should cope with ever-growing IoT traffic. This work proposes a S-ALOHA system with time offsets (TOs), where one slot consists of $K$ TOs and one packet transmission time. The length of the overall TOs is a fraction of a packet transmission time. In the system users (re)transmit to the boundary of a TO randomly selected. This enables the base station (BS) to inform the users of who transmits the first and the last packets in the slot with collision so that the users transmitting the first and last packet can retransmit successfully in the following two slots, respectively. Our throughput analysis compared to simulations shows that adopting even with three and four TOs surpasses the throughput limit of S-ALOHA system without TOs. Additionally, we propose two Bayesian-optimized backoff algorithms for S-ALOHA system with TOs, with which users can apply throughput-optimal (re)transmission probability or uniform backoff window even in unsaturated traffic scenarios. Numerical results demonstrate that the proposed backoff algorithms can achieve the throughput close to an ideal system and drastically reduce the access delay compared to S-ALOHA system.   

Proliferation of the urban Internet-of-Things (IoTs) for smart cities has fuelled massive amounts of data over wireless cellular networks. Random access (RA) system of wireless cellular networks, e.g., 5G New Radio (NR), based on S-ALOHA system should cope with ever-growing IoT traffic. This work proposes S-ALOHA system with time offsets (TOs), where one slot consists of K TOs and one packet transmission time. The length of the overall TOs is a fraction of a packet transmission time. In the system users (re)transmit to the boundary of a TO randomly selected. This enables the base station (BS) to inform the users of who transmits the first and the last packets in the slot with collision so that the two users can retransmit successfully in the following two slots respectively. Our throughput analysis compared to simulations shows that adopting even with three and four TOs surpasses the throughput limit of S-ALOHA system without TOs. Additionally, we propose two Bayesian-optimized backoff algorithms for S-ALOHA system with TOs, with which users can apply throughput-optimal (re)transmission probability or uniform backoff window even in unsaturated traffic scenarios. Numerical results demonstrate that the proposed backoff algorithms can achieve the throughput close to an ideal system and drastically reduce the access delay compared to S-ALOHA system.

%examine  the effect of nonidentical values of user's parameters on each user's throughput. In particular, the throughput region is obtained by formulating a multiobjective optimization problem (MOOP) for  two users' throughput and compared with that of slotted ALOHA systems.
  %
   %and derive the Laplace Stieltjes transform (LST) of the probability density function (pdf) of the access delay, from which the moments of the access delay can be obtained. We obtain  In $N$-user system,  
\end{abstract}


%\IEEEdisplaynontitleabstractindextext
\IEEEpeerreviewmaketitle

%\IEEEraisesectionheading{}
\section*{Nomenclature}\label{sec:nom}
\addcontentsline{toc}{section}{Nomenclature}
\begin{IEEEdescription}[\IEEEusemathlabelsep\IEEEsetlabelwidth{$C_i\quad$}]
	\item[$C_i$] The event of type-$i$ collision for $i=1,2$.
	\item[$S$] The event of success by a single packet transmission in a slot
	\item[$S_{i,c}$] The event of type-$i$ success under type-1 collision for $i=0,1,2$. 
		\item[$S_{i}$] A successful packet transmission in the  first (or second)  following slot upon type-1 collision, irrespective of the transmission result in the  second (or first) following slot
	\item[$\omega$] The event of a three-slot collision  under type-1 collision
		\item[$I$] The event of an idle slot
			\item[$\mathbb{C}$] The event of collision in a slot, either type-1 or 2
				\item[$\alpha$] The length of each time offset (TO)
								\item[$K$] The number of TOs in a slot
									\item[$T$] The length of a packet transmission
									\item[$T_s$] The length of one slot: $T_s\triangleq(K-1)\alpha+T$
									\item [$\gamma$] A ratio of a packet transmission time to a slot length: $T/T_s$
									\item[$\mathcal{B}_i^n(p)$] Binomial distribution: ${n\choose i}p^i(1-p)^{n-i}$
										\item[$\Phi_n(\nu)$] Poisson distribution with mean $\nu$: $\frac{\nu^n}{n!}e^{-\nu}$
\end{IEEEdescription}

\section{Introduction}\label{sec:Introduction}
\subsection{Motivations}
As
various heterogeneous wireless communication networks such as Long-Term Evolution (LTE), IEEE 802.11, Zigbee, and Bluetooth have seamlessly interworked around a decade ago, access to the Internet  became ubiquitous. This has also enabled remote objects, e.g., smart sensors, to collect and exchange data with minimal human intervention, which is often referred to as the Internet-of-Things (IoTs) \cite{Fuq}. Collected data is analyzed with big data analytics and artificial intelligence (AI) so as to find hidden behavior and correlated pattern, or make optimal decisions and management even in real-time. As IoT applications have grown huge and diverse for smart cities, autonomous vehicle, health, logistics and supply chain optimization, etc., the number of IoT devices and the volume of IoT data have explosively increased. It might not be difficult to forecast that IoT applications would be more concentrated in densely populated  urban areas due to economic reasons. In order to cope with massive amounts of IoT data in urban areas, wireless cellular networks, e.g., LTE, the fifth generation (5G) New Radio (NR), would be well-suited due to their ubiquitous deployments worldwide. 

As random access (RA) system for cellular networks, slotted ALOHA (S-ALOHA) systems have long been adopted from the second generation (2G)  network, i.e., Global system for mobile communication (GSM) to LTE and 5G NR. Although the maximum throughput of S-ALOHA system is limited up to $e^{-1}\approx  0.3679$ (packets/slot) for a large population size, it has been ever used in LTE and even further 5G NR, by making use of orthogonal RA preambles.  The maximum throughput can be scaled by a factor of the number of  RA preambles  deployed  in LTE and 5G NR. However,  the throughput per RA preamble is still limited by $0.3679$. Another limiting factor is  that RA protocol in LTE and 5G NR  is based on a four-way handshaking procedure, where data channel request after RA preamble transmission should be granted before data transmission. RA systems in LTE and 5G NR thus could be a bottleneck due to massive accesses from an unprecedentedly growing number of IoT devices, while data freshness is ever more required. It is essential to develop  a new RA system for the next generation wireless access system, which is simple enough to be implemented, while yielding higher throughput  to accommodate a surge of IoT traffic.  This work proposes a new S-ALOHA system employing time-offsets (TOs) in slots. 

\subsection{Related Works}
Before introducing what schemes are devised for S-ALOHA system to cope with massive IoT traffic in the literature, it is noteworthy that the duty cycle and data rate of IoT devices vary according to their applications \cite{Siv}. Devices might gather data for some time period without transmission and will transmit in a batch later. Traffic from innumerable IoT devices can be mixed in time and is transmitted to gain access to S-ALOHA system, competing with  human mobile devices. 

As prior work to deal with them, access class barring (ACB) schemes have been applied to S-ALOHA system of LTE in \cite{Cheng,Duan,Cheng2,Jin}. Applications are classified into 10 classes in order of urgency. Then, access from devices belonging to some classes may be temporarily blocked such that traffic of higher classes can be protected from access of traffic of lower classes. While ACB schemes control burstiness of IoT traffic, they can not overcome the throughput limit of S-ALOHA system. It is notable that ACB scheme can be used with any RA systems introduced below, since it is a burstiness regulator.

In order to achieve throughput higher than $0.3679$, multipacket reception (MPR) channels have been proposed in \cite{Ghez,Baio, Naware}: While at least two packets are involved in a collision, none of them is decoded in S-ALOHA without MPR capability. However, in MPR-capable S-ALOHA system, with the help of  some advanced signal processing techniques, e.g., successive interference cancellation (SIC), or multi-input multi-output (MIMO) antenna in physical layer, one or more packets in a collision can be recovered. How many packets can be decoded in a collision can lead to  a higher throughput. Throughput gain with MPR channel comes at the expense of additional resources and computational complexity in physical layer, e.g., codes, transmission power and/or SIC. From medium access control (MAC) layer's perspective, it can be  abstracted into $k$-out-of-$m$ channels; that is, $k$ successes among $m$ simultaneous accesses in a slot. This abstraction includes RA systems with orthogonal multichannel, e.g., frequency or code; thus, LTE with orthogonal RA preambles can be also considered a kind of MPR channel. 

% can be successfully recovered  in MPR channel,

Another effort of getting a higher throughput of S-ALOHA system is to exploit retransmission diversity with SIC in \cite{Casini,Liva,Pao, Jeon}. Even with some variants, basically it works as follows: Multiple copies of a packet are transmitted over different slots, in which those of other users' packet \emph{can} be transmitted. Each copy of a packet contains the information about where other copies of the same packet are transmitted. If one of them is successfully decoded in a slot, the base station (BS) with SIC utilizes it to decode other packets transmitted with it in other slots. In practice, it remains to be seen how SIC works in an efficient and fast way over multiple slots. 

As an alternative to S-ALOHA system, distributed queueing random access (DQRA) scheme \cite{Xu,Alon} has been applied to LTE in \cite{Laya,Ahn}. 
In Xu's original DQRA scheme \cite{Xu}, a slot is divided into mini-slots and one data transmission part. Instead of transmitting a data packet directly, users first transmit a request to mini-slots based on tree (or splitting) algorithm. Those users who have transmitted a request successfully are then queued in a logical queue waiting for data packet transmission. LTE RA system separates RA preamble (request) and data transmissions into physical RA channel (PRACH) and physical uplink shared channel (PUSCH). However, users transmit RA preambles to PRACH based on S-ALOHA system. Accordingly, it seems that the use of DQRA system is  to replace S-ALOHA system with a tree algorithm for RA preamble transmission. 

In general, tree algorithms yield a higher throughput than S-ALOHA system: The best algorithm known so far,  i.e., first-come and first-serve (FCFS) algorithm \cite{Bert}, achieves $0.4871$-throughput (packets/slot). High throughput might result from that an initial group of users in collision keeps being divided into smaller groups  during a contention resolution period (CRP), for which new accesses are blocked. Notice that even in S-ALOHA system, throughput can be improved if the number of contending users is restricted to a small number. While the sequence of retransmissions from the small backlogged groups forms a tree,  idle slots in the CRP period are expected to be avoided as much as possible.  A  random splitting algorithm has been introduced to improve the adaptability of FCFS to  network dynamics \cite{Toor_mobihoc}.   To get even higher throughput, tree algorithms have been combined with MPR channels in \cite{Wang, Waqas,Gau,Yim,Gore}. However, compared to S-ALOHA system, two drawbacks are expected for implementation. Firstly, backlogged and non-backlogged users should keep looking for the beginning and the end of the CRP, which results in inefficiency of energy consumption for IoT devices. Secondly, the retransmission sequence from a large number of the small groups should be robustly maintained in order not to lose synchronization in practice. Otherwise, the system may be prone to collapse.

For IoT devices to transmit a small-sized data sporadically, the four-step hand-shaking protocol in LTE seems heavy and has protocol-wise latency. To tailor it for such IoT traffic, two-step RA systems have been studied in \cite{Huang,Jun1,Kim,Choi}. Two-step RA protocol in short focuses on where and how to transmit a RA preamble and data packet simultaneously. 

In comparison with the previous work,  our proposed system adopts $K$ TOs in each slot, still making use of S-ALOHA system. In contrast with mini-slots of DQRA system, which are used for requesting packet transmissions, the overall length of TOs in a slot is much smaller than one packet transmission time for transmission efficiency. In our system,  transmissions of requests  and data packets are not separated: Users' (data) packets  can be transmitted directly at one of TOs in a slot so that those packets can be overlapped in a slot. Because of the direct transmission of \emph{data} packet with getting access granted, the proposed system might eliminate latency in the four-step handshaking protocol in LTE system. Furthermore, we shall see that the achievable throughput of the proposed system far exceeds $0.3679$, while any form of MPR channels is not used. This implies that 
the proposed scheme could be flexibly implemented in MAC layer, requiring no additional resources in physical layer. As a rough comparison with tree algorithms, the proposed scheme consumes only two slots after some particular collisions. If the collision followed by two slots in the proposed system could be called CRP, it takes three slots at maximum. In order to stabilize the proposed system, we propose two Bayesian backoff algorithms. In one algorithm with uniform window, backlogged users do not need to monitor the broadcast message from the BS continuously as in FCFS algorithm. Yet, the proposed system can achieve higher throughputs than $0.4871$ of FCFS algorithm. Although ACB scheme can be applied to the proposed system, it is out of our scope. %, 

\subsection{Contributions}
Our contribution can be summarized as follows.
\begin{itemize}
	\item This work designs a new S-ALOHA system by introducing TOs to slots: When users have a packet to send, they (re)transmit their packet at one of TOs randomly in a slot. Upon a collision, by observing the transmissions at these TOs, the BS enables the users to realize who transmits the first and the last packets such that they retransmit immediately in the  two following slots, respectively.  In order to demonstrate the performance of the proposed system, we carry out throughput analysis for saturated systems and validate it with simulations. We show how the duration of a TO and the number of TOs affect the performance and can be optimally chosen. 
	\item 
	We further develop two Bayesian backoff algorithms to be used in practice for the proposed system: In one system, the BS broadcasts a throughput-optimal (re)transmission probability, and throughput-optimal uniform window size in the other. The advantage of the algorithm with uniform window over retransmission probability is less energy consumption and is easy to implement in terms of monitoring BS's broadcast messages. Both algorithms are based on Bayesian optimization in the sense that building \emph{a prior} belief over the unknown number of backlogged users who have packets to transmit, the algorithm updates the prior from the posterior distribution of the backlogged users given the channel outcome recursively over time. Finally, we show that the system with a fixed retransmission probability is always unstable, while the proposed Bayesian backoff algorithm can stabilize the system. 
	\item Extensive simulations are conducted for validation; the proposed ALOHA with TOs and backoff algorithms are compared against S-ALOHA system without TOs and FCFS algorithm under Poisson traffic as well as Beta-distributed traffic. %This work demonstrates the performance of the proposed system with traffic model of Poisson process as well as that of Beta distribution for massive IoT traffic. }
	
\end{itemize}
\subsection{Organization}
The organization of this work is as follows: In Section \ref{sys}, S-ALOHA system with TOs is introduced. Section \ref{sec:ana} provides the throughput analysis  and how the Bayesian backoff algorithms work and are derived. Numerical results are discussed in Section \ref{sec:num} and concluding remarks are finally given in Section \ref{sec:con}.

\section{S-ALOHA System with time offsets}\label{sys}

\begin{figure}[pt]
	\centering
	\includegraphics[width=7.cm, height=3cm]{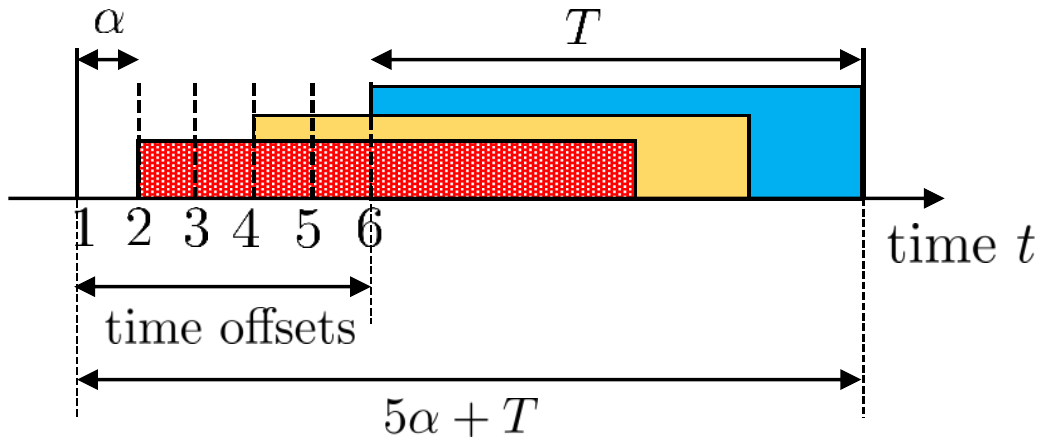}
	\caption{Slot structure $T_s$ with six TOs ($K=6$).}\vspace{-0.2cm}
	\label{fig:figm1}
\end{figure}

In S-ALOHA system, a BS is located at the center of its circular coverage area, where users are randomly scattered and wirelessly connected with the BS. Time is divided into slots of a constant length, $T_s$ (sec) long. Furthermore, one slot  consists of $K$ time offsets (TOs) plus one packet transmission time $T$ as shown in Fig. \ref{fig:figm1}, where six TOs are shown and three rectangles indicate three packets in transmission. The length of each TO is a constant denoted by $\alpha$ and a packet transmission can occur at one of TOs. Therefore, one slot length $T_s$ is equal to $T_s=\alpha(K-1)+T$.  The length of the overall TOs is assumed to be smaller than one packet transmission time, i.e., $\alpha (K-1)< T$ for transmission efficiency of the time slot. 

We define that each slot has two states, say \emph{open} and \emph{closed}, which is broadcast by the BS with a downlink feedback message.  Users freely transmit their packets to an open slot, not a closed slot. In sending their packet to an open slot, the  users go through a Bernoulli trial with (re)transmission probability $p$; that is, each user draws a random number in the interval between 0 and 1. If it  is less than $p$, the user transmits; otherwise, no transmission. The users repeat this trial in every open slot. Upon (re)transmission, the user chooses one of TOs randomly and transmits the packet at its boundary. Note that if a packet is transmitted at the boundary of the last TO, e.g., the 6-th TO shown in Fig. \ref{fig:figm1},  the packet transmission time ends at the slot boundary such that it can not interfere with the transmissions in the next slot. 

\begin{figure*}[pt]
	\centering
	\includegraphics[width=0.9 \textwidth]{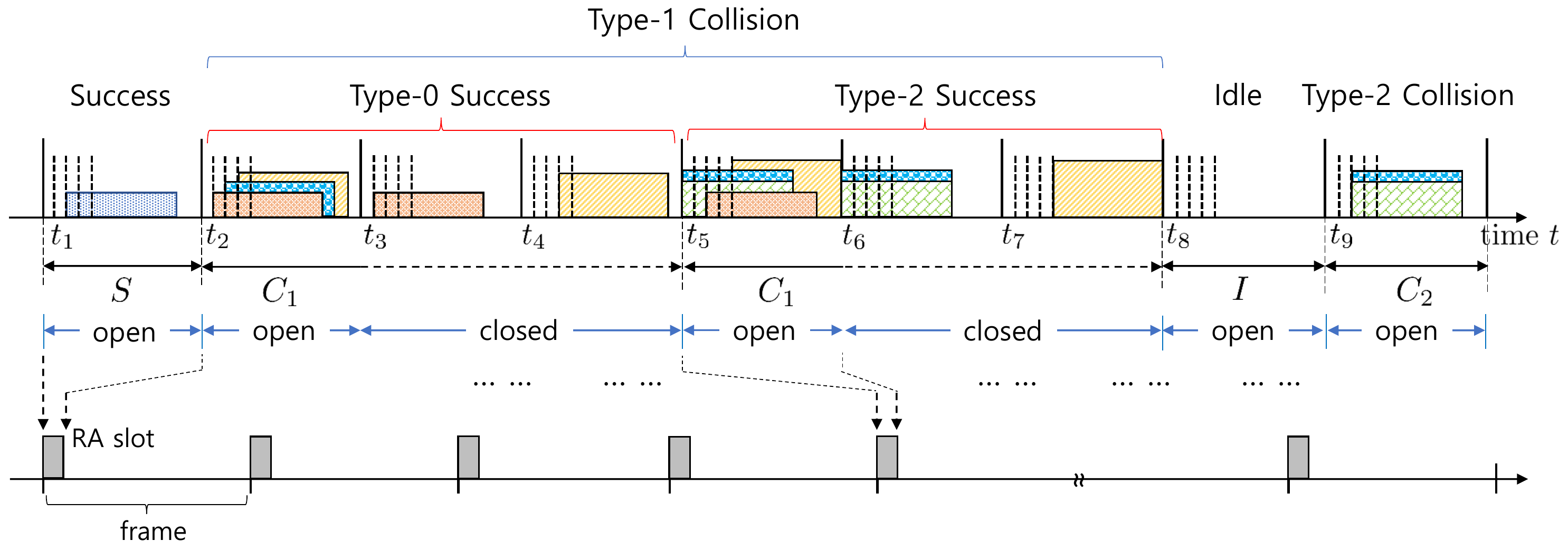}
	\caption{Various channel outcomes in S-ALOHA system with TO.}
	\label{fig:figm2}
\end{figure*}

As in S-ALOHA system without TOs, if only one packet is transmitted in a (open) slot, the transmission is declared successful. This channel outcome is informed by the BS with a feedback message over the downlink.  A slot without a packet transmission is called idle. Fig. \ref{fig:figm2} illustrates various channel outcomes. Success and idle open slots can be observed at slots $t_1$ and $t_8$, respectively.  In these two cases, the state of the next slot is always open. The slots at $t_4$ and $t_7$ are success  but closed. This shall be explained soon.

% first   When a slot is open, users transmit their packet (at one of time offsets in the slot) according to Bernoulli trial with probability $p$
When more than one packets are transmitted in an open slot as shown in Fig. \ref{fig:figm1} or at slots $t_2$ and $t_5$ in Fig. \ref{fig:figm2}, a collision occurs. In these cases, the BS extracts two pieces of information; that is, two TOs, where the first packet and the last packet transmissions take place in a slot.  The BS can obtain the information on the aforementioned two TOs by observing at which TO   the first packet is transmitted in the slot and where the transmission ends. To this end, we assume that the users use a transmit power control scheme so that the received power of the packet is a predefined value, i.e., decoding threshold\footnote{This is often called channel inversion \cite{gold}; this is enabled in various ways. For example, in time-division duplex (TDD) mode, the BS broadcasts a reference signal and specifies its transmit power. Then, the users estimate the uplink channel gain from the downlink reference signal. This facilitates that the BS decodes a single packet transmission and finds the beginning and the end of a transmission.} To detect the beginning and end of a transmission, i.e., collision, in a slot, we can consider two methods \cite{Hei,Tan,Alou}. The first one is that the BS can make use of an energy detection used to detect a collision as in IEEE 802.11 wireless local area networks (WLANs). The access point (AP) and the users sense the channel using energy detection and detect the beginning and end of a collision. Then, they resume the backoff count in carrier sense multiple access (CSMA) right after collision. In the proposed system, only the BS can find the beginning and end of a collision using energy detection. Notice that since the length of packet is fixed and known in S-ALOHA, subtracting one packet length from the end of a transmission gives the TO, where the last packet is transmitted. A second method is that a preamble and a postamble can be appended to each packet. The BS can use a sliding window technique with the preamble and postamble to find the beginning of the first packet and the end of the last packet in the transmission. So long as the packets are aligned at TOs, even if some packets are superposed at the TO that the transmission begins or ends, correlating the preamble or postamble with incoming packets can yield some peak energy, at which the TOs are identified.  Once detecting the end of a collision,  the BS  can figure out the TO, where the last packet transmission begins, by subtracting one packet transmission time from the end of the overall transmission.  %It is also notable that the packet transmission at each TO boundary can be slightly relaxed as shown Fig. , where the peak energy by correlating the preamble and postamble can be detected between the $i$-th and the $(i+1)$-th TO. }

We define two types of collisions in open slot. Since it is possible that all the packets in a collision could be transmitted at the same TO, the collision that at least one packet is transmitted at a different TO is defined as type-1 collision; otherwise, type-2 collision. Type-1 collision is  found at slots $t_2$, and $t_5$ in Fig. \ref{fig:figm2}, while type-2 collision at slot $t_9$, i.e., all packets transmitted at the same TO. Although the collision at slot $t_6$ looks like type-2 collision, it is differently treated as it occurs within a closed slot, which is also explained later. 

Let us describe the retransmission rules upon each type of collisions. Upon type-1 collision, just before slot $t_3$, the BS broadcasts the information for the two TOs and the occurrence of type-1 collision, and additionally announces that the next two slots are closed. Thus, slots $t_3$, $t_4$, $t_6$, and $t_7$ are \emph{closed} slot due to the collision at slots $t_2$ and $t_5$.  Since the users involved in type-1 collision now know which TO they have chosen, the information for the two TOs enables them to realize whether their packet is either the first transmission, or the last one in the slot. For instance, suppose that the BS informs TOs 2 and 6 in Fig. 1 right before the next slot. %Users that own the packet transmitted at TOs 2 and 5 
When the users realize that their packet is the first one transmitted in the slot, they transmit their packet at any TO in the following (closed) slot. Furthermore, those who find their packet the last one in the slot will transmit their packet at any TO in the second following (closed) slot. Fig. \ref{fig:figm2} shows that the packets transmitted first and last in slot $t_2$ are retransmitted at slot $t_3$ and $t_4$, respectively. Therefore, in the two slots following the slot of type-1 collision, four cases are expected to observe. \emph{Firstly}, if one packet is transmitted at each TO, where the first packet transmission and the last one occurs, respectively, these two packets are transmitted in the  first following slot and the  second following slot each. This results in two successful packet transmissions in the following two (closed) slots, e.g., slots $t_3$ and $t_4$. This is defined as type-0 success under type-1 collision. \emph{Secondly}, suppose that only one packet is transmitted earliest at a TO, but multiple packet transmissions occur last at a TO in the same slot. Then, it can be expected that one packet is successfully transmitted in the  first following slot, while a collision is found in the  second following slot. \emph{Thirdly}, as opposed to the second case, there can be multiple packet transmission earliest at a TO, while one packet is transmitted last at a TO in the same slot. This can be found at slot $t_5$ in Fig. \ref{fig:figm2}. Accordingly, we expect a collision in the  first following slot, i.e., slot $t_6$, and a success in the second following slot, i.e., slot $t_7$. The cases with a single success above are called type-1 and type-2 success under type-1 collision, respectively.   Unfortunately, as the fourth case, multiple packets are transmitted earliest and last, respectively. If so, no packets (re)transmitted in the two successive slots are successfully received. This is particularly called three-slot collision under type-1 collision. Note that the state of the next slot after two closed slots due to type-1 collision will be always open. For two closed slots, no new packet transmissions are allowed.
\begin{table*}[pt]
		\caption{Summary of channel outcomes\label{tab:tab0}}\vspace{-0.3cm}
		\begin{center}
			\begin{tabular}{c|c|c|c|c|c}
				\hline
				\multicolumn{2}{c|}{Outcome of an open slot} & State of next slot(s)&  $\#$ of successes per slot& 
	 $\#$ of accesses per slot	& 		Additional info.	\\\hline
				\multicolumn{2}{c|}{Success} & Open & $1/1$  &1&none
				\\\hline
			\multirow{4}{*}{Type-1 collision}&Type-0 Success& \multirow{4}{*}{Closed for next two slots} & $2/3$ & $2$ & \multirow{4}{*}{index of two TOs}	 \\\cline{2-2}\cline{4-5}
				&Type-1 Success& 	& $1/3$ & $\geq 3$& \\\cline{2-2}\cline{4-5}
					&Type-2 Success& 	& $1/3$ & $\geq 3$&	 \\\cline{2-2}\cline{4-5}
				& Three-slot collision&  & $0/3$ & $\geq 4$ \\\cline{1-6}
							\multicolumn{2}{c|}{Type-2 collision} & Open &  $0/1$& $\geq 2$ (at the same TO) &none\\
				\hline
			\end{tabular}
			\\\vspace{0.1cm}
			\begin{flushleft}
			\end{flushleft}
		\end{center}\end{table*}
Let us move on to type-2 collision, where all the packets in a slot have been transmitted at the same TO.  The BS can find this when it finds a transmission at a TO, but its receive power is greater than a predefined value. Upon this collision, the BS  can inform type-2 collision, e.g., slot $t_9$ in Fig. \ref{fig:figm2} and sets the state of the next slot to be open. This enables any users to access with probability $p$ again.  Table \ref{tab:tab0} summarizes the overhead of the feedback information can be two bits to represent the channel outcome, i.e., success, type-1 or type-2 collision, and index of two TOs, i.e., $\log_2 K$ bits; thus, the total bits that we need are $2+\log_2 K$ bits.  

We make the following assumptions: First, the feedback message is immediately available.  In practice, this assumption could be justified when a RA channel appears periodically in a slot of the frame in Fig. \ref{fig:figm2} such that there are enough slots for the BS to inform the feedback message between two RA channels. It should be noted that the slots along the time axis in Fig. \ref{fig:figm2} are drawn in a generic form.  Second, we assume a strong channel coding scheme. Thus, the feedback message broadcast by the BS over the downlink  is delivered to the users without an error, while the packets transmitted by users over uplink can not be decoded only due to collisions. 

%\section*{Nomenclature}
%\addcontentsline{toc}{section}{Nomenclature}
%\begin{IEEEdescription}[\IEEEsetlabelwidth{$V_1,V_2,$}]
%	\item[\smash{\begin{IEEEeqnarraybox*}[][t]{l}
%			V_1,V_2,\\
%			\hphantom{V_1,{}}V_3
%	\end{IEEEeqnarraybox*}}] Three-phase PWM output line voltages.\\
%	\mbox{}
%	\item[$\theta$] Rotor angle (in ``electrical degrees'').
%	\item[$\omega$] Rotor (electrical) speed, corresponding to the time
%	derivative of $\theta$.
%\end{IEEEdescription}

\begin{figure}[pt]  \centering
%	\subfigure[Case 1.]{	
		\includegraphics[width=7.cm, height=3.1cm]{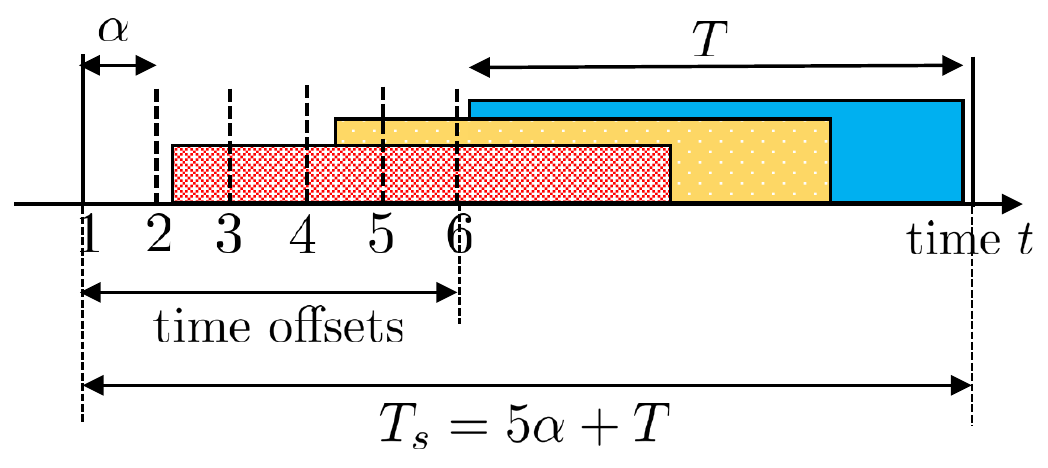}	\label{fig:fig222a}%}
%	\subfigure[Case 2.]{	
%		\includegraphics[width=8.cm, height=3.6cm]{slot_loose2}	\label{fig:fig222b}}
	\caption{The last part of a slot larger than one packet transmission time.}		\label{fig:fig222}
\end{figure}

The proposed S-ALOHA with TOs is not necessarily intended for the current LTE and 5G NR. For practical implementation, a high level of synchronization and slot structure with TOs in Fig. \ref{fig:figm2} would be prerequisites to apply the proposed system to LTE and 5G NR, or the next generation wireless network.  We assume that the users learn uplink synchronization through initial and periodic cell association processes.  It is notable that feasibility of a high level of synchronization and the use of TOs can be found in the use of mini-slots in 5G NR \cite{tr912}. More specifically, short packet transmissions over mini-slots recently have been considered in a slot. Additionally,  as in Fig. \ref{fig:fig222}, the interval between each TO and the last part for one packet transmission time in a slot are made slightly larger to absorb coarse synchronization at the boundary of TOs. A more detailed discussion may be beyond our scope. So, we leave for future work.
%%%%%%%%%%%%%%%%%%%%%%%%%%%%%%%%%%%%%%%%%%%%%%%%%%%%%%%%%%%%%%%%%%%%%%%%%%%%%%%%%%%%%%%%%%%%%%%%%%%%%%%%%%%%%%%%%%%%%%%%%%%%%%%%%%%%%%%%%%%%
% One slot consists of 14 orthogonal frequency division multiplexing (OFDM) symbols  in 5G NR, where mini-slot can  take 2, 4, or 7 OFDM symbols. A packet can be transmitted at any OFDM symbol in a slot, which can last as many symbols as mini-slots. 
%	Although this packet transmission is not intended for RA in 5G NR, we might assume that the slot structure in Fig. \ref{fig:figm1} with such a high level of synchronization can be placed in a frame consisting of 10 slots. Then, the users can transmit their packet to the slot as in the proposed system. On the other hand, if the next generation  network supports a new slot structure, it would be possible to design the TOs much shorter than the OFDM symbol discussed above.  It is notable that the BS just needs to distinguish the beginning of each TO without decoding the information bits in each TO upon a collision.
%%%%%%%%%%%%%%%%%%%%%%%%%%%%%%%%%%%%%%%%%%%%%%%%%%%%%%%%%%%%%%%%%%%%%%%%%%%%%%%%%%%%%%%%%%%%%%%%%%%%%%%%%%%%%%%%%%%%%%%%%%%%%%%%%%%%%%%%%%%

\section{Analysis and Algorithm Design}\label{sec:ana}
\subsection{Throughput Analysis}
\begin{Theorem}\label{Th:Th1}Suppose the system with $K$ TOs, where each of $n$ users always has a packet to transmit, i.e., \emph{saturated} user. 	
	Throughput of  this system  (packets/$T$) can be obtained as 
\begin{equation}\label{eq:eq61}
	\tau_{n}(p)=\gamma \cdot  \frac{\mathcal{B}_{1}^{n}(p)+2 \displaystyle\sum_{i=2}^{n} \sum_{j=1}^{K-1} \frac{i \displaystyle\mathcal{B}_{i}^{n}(p)}{K^{i}}(K-j)^{i-1}}{3-2\left(\sum_{i=0}^1\mathcal{B}_{i}^{n}(p)+\sum_{i=2}^{n} \frac{\displaystyle\mathcal{B}_{i}^{n}(p)}{K^{i-1}}\right)},
\end{equation} 
where $\gamma=T/T_s$, and $T_s\triangleq (K-1)\alpha+T$.
\begin{IEEEproof} 
See Appendix \ref{App1}.
\end{IEEEproof}
\end{Theorem}

%finally we can get $\text{Pr}\left[  S_{2,\varphi}\right]$:
%\begin{align}
%	\Pr\left[  S_{2,\varphi} \right] =\sum_{i=2}^n\mathcal{B}_{i}^{n}(p)\Pr\left[  S_{2,\varphi}\mid i \right].
%\end{align}
%We can find  $\text{Pr}\left[ CS_2 \right] =\text{Pr}\left[ CS_1 \right]$.
%	In fact, we shall see 
%\begin{align}
%	\Pr[S_{1,c}]=\Pr[S_{2,c}]
%\end{align}
%\begin{align}
%	\Pr[C_1]= \sum_{i=0}^{2}\Pr[S_{i,c}] + \Pr[\omega]
%\end{align}
%So, we can get throughput:
%\begin{equation}
%	\begin{aligned}
%		S&=\frac{\mathbb{E}[R]}{\mathbb{E}[\mathcal{Z}]}\frac{\Pr\left[ S \right] +2\Pr\left[ S_{0,c}\right] + \sum_{i=1}^2\Pr\left[S_{i,c} \right] }{\left[ \text{Pr}\left[ I \right] +\text{Pr}\left[ S \right] +\text{Pr}\left[ C_1 \right] +3\left( \text{Pr}\left[ C \right] -\text{Pr}\left[ C_1 \right] \right) \right] \left( T+\alpha \right)}\\
%		&=\frac{\text{Pr}\left[ S \right] +\text{2Pr}\left[ CS_1 \right]}{\left[ 3-2\left( \text{Pr}\left[ I \right] +\text{Pr}\left[ S \right] +\text{Pr}\left( C_1 \right] \right) \right] \left( T+\alpha \right)}
%	\end{aligned}
%\end{equation}
\begin{Corollary}\label{Cr:Cr1} As $n\rightarrow\infty$, $\tau_n$ is approximated as 
%	\begin{align}
%		\tau_{n\rightarrow \infty}=\frac{n p e^{-n p}+\frac{2 n p}{K}\left[\sum_{j=1}^{K-1} \left(e^{-\frac{n p}{K}}\right)^j-(K-1) e^{-n p}\right]}{\left[3-2 e^{-n p}\left(K e^{\frac{n p}{K}}-K+1\right)\right] T_{s}},
%	\end{align}
	\begin{align}\label{eq:eq22}
	\tau_{\infty}&\approx \gamma \cdot \frac{\eta e^{-\eta }+\frac{2 \eta}{K}\Bigg[\frac{e^{-\frac{\eta}{K}}-e^{-\eta}}{1-e^{-\frac{\eta}{K}}}-(K-1) e^{-\eta}\Bigg]}{3-2 e^{-\eta }\left(K \left(e^{\frac{\eta}{K}}-1\right)+1\right)},
\end{align}
where $\eta=np$. 
\begin{IEEEproof} See Appendix \ref{App2}.
	\end{IEEEproof}
\end{Corollary}
\begin{Corollary}
Average access delay $\mathbb{E}\left[ D|n \right] $ is
\begin{equation}
	\mathbb{E}\left[ D|n \right] =\frac{n}{\tau_n}.
\end{equation}
\begin{IEEEproof} This can be proved by Little's result \cite{Bert}, i.e., the number of backlogged users $n$ in the system equals to  $\tau_n\mathbb{E}[D|n]$.
	\end{IEEEproof}
\end{Corollary}

	If the system has infinitely many TOs and infinitesimally short length of TO, i.e., $K\rightarrow\infty$ and $\alpha\rightarrow 0$, we can expect that the system achieves the upper-bound of throughput. \begin{Proposition}\label{Pr:Pr1} The throughput upper-bound of S-ALOHA with TOs is expressed as
	\begin{align}\label{eq:tau_o_infty}
		\tau^o_\infty =\frac{2-(2+np)e^{-np}}{3-2e^{-np}(1+np) }.
	\end{align}
\begin{IEEEproof} See Appendix \ref{App3}.
	\end{IEEEproof}
\end{Proposition}
In Proposition \ref{Pr:Pr1}, the throughput upper-bound is expressed as a function of $np$. Let us examine the maximum throughput, denoted by $\hat{\tau}_{\infty}^o$, for $np$ in the following corollary. For a large population, the system can not provide more than this due to $\alpha\rightarrow 0$ and $K\rightarrow \infty$. 
\begin{Corollary}\label{Cr:Cr3} The maximum of $\tau_\infty^o$ with respect to $np$ is 
	\begin{align}
		\hat{\tau}_{\infty}^o=0.673.
	\end{align}
	\begin{IEEEproof} To find the optimal $np^*$, which maximizes the $\tau_\infty^o$, we calculate the $np$ that makes $\frac{d\tau_\infty^o}{d(np)}=0$, that is $2e^{-np}+np-3=0$. we can get $np^*=2.89$. Plugging this into  \eqref{eq:tau_o_infty}, we can get $\hat{\tau}_{\infty}^o=0.673$.
	\end{IEEEproof}
\end{Corollary}
Proposition \ref{Pr:Pr1} can be also obtained from Corollary \ref{Cr:Cr1}. Letting $K\rightarrow \infty$, we can find $e^{-\frac{\eta}{K}}=1$, $\frac{K-1}{K}=1$ and write \eqref{eq:eq22}  as
\begin{equation}
\begin{aligned}
\tau_{\infty} & \approx \frac{\eta e^{-\eta}+2 \eta\left[\frac{e^{-\frac{\eta}{K}}-e^{-\eta}}{K(1-e^{-\frac{\eta}{K}})}-\frac{K-1}{K} e^{-\eta}\right]}{3-2 e^{-\eta}\left(K\left(e^{\frac{\eta}{K}}-1\right)+1\right)}\\
&=\frac{\eta e^{-\eta}+2 \eta\left(\frac{1-e^{-\eta}}{\eta}-e^{-\eta}\right)}{3-2 e^{-\eta}\left(\eta+1\right)}\\
&=\frac{2-(2+\eta)e^{-\eta}}{3-2e^{-\eta}(1+\eta)},
\end{aligned}
\end{equation}
where L'Hopital's rule has been used. This is identical to \eqref{eq:tau_o_infty}.
%So when $\gamma=1$ and $K=\infty$,
%Plugging $\eta=np$, we can get \eqref{eq:tau_o_infty}.

%%%%%%%%%%%%%%%%%%%%%%%%%%%%%%%%%%%%%%%%%%%%%%%%%%%%%%%%%%%%%%%%%%%%%%%%%%%%%%%%%%%
\begin{algorithm} [pt]
	\caption{Proposed Bayesian backoff algorithm}
	\begin{algorithmic}[1] 
		\label{algo1}
		\STATE Initialize $\lambda_0=0$, $\nu_0=1$, $p^*_1=1$,  $L_0=1$, and do this every open slot $t=1,2,...$.
		\STATE $\lambda_t=\theta \lambda_{t-1}+(1-\theta)\frac{(\mathbb{I}(S)+s\cdot\mathbb{I}(C_1))}{L_{t-1}}$.
		\IF {the current slot is idle or success }
		\STATE $\nu_t=\max(\nu_{t-1}-\kappa,0)$ and $L_t = 1$
		\ELSIF {the current slot is type-1 collision}
		\STATE $\nu_t=\max(\nu_{t-1}+\frac{\kappa}{e^{\kappa}-\kappa-1},2)-s$ and $L_t=3$
		\ELSE  
		\STATE $\nu_t=\max(\nu_{t-1}+\frac{\kappa}{e^{\kappa}-\kappa-1},2)$ and $L_t=1$.
		\ENDIF
		\STATE $\nu_t \leftarrow \nu_t +   \lambda_t L_t $.
		\STATE Broadcast $p^*_{t+1}=\min(\kappa/\nu_t,1)$ for the next open slot.
	\end{algorithmic}
\end{algorithm}
%%%%%%%%%%%%%%%%%%%%%%%%%%%%%%%%%%%%%%%%%%%%%%%%%%%%%%%%%%%%%%%%%%%%%%%%%%%%%%%%%%%

\subsection{Algorithm Design}
In practice, the number of users or IoT devices that have a new packet to send and then join the backlog is changing randomly in every slot.  This randomness also varies over different traffic models. Meanwhile, other backlogged users have kept retransmitting with a (re)transmission probability, say $p_t$ at an open slot $t$. Depending on $p_t$, the number of backlogged users, i.e., backlog size, can decrease over time, or vice versa. In this section, we develop two backoff algorithms to control $p_t$: The first one is that the BS broadcasts  a throughput-optimal (re)transmission probability $p_t$ so that backlogged users can use it for Bernoulli trial for (re)transmission. If some of them do not retransmit with probability $1-p_t$, they have to get a new $p_{t+1}$ given in the next broadcast message. To save more energy that the users may consume for monitoring BS's broadcast message, we will propose a modified version of this algorithm in the next section.

%gain, since access to closed slots is limited, the proposed algorithm controls access only to open slots. 

Before deriving each step of the proposed algorithm, we first introduce how it works. In Algorithm \ref{algo1},  $\lambda_t$ (packets/slot) is the BS's estimation on the mean rate of new packet arrivals at slot $t$, and $\mathbb{I}(x)$ denotes an indicator function, which takes one if $x$ is true; otherwise, zero. It is notable that this kind of mean estimation does not depend on the traffic models. In addition, $s$ indicates the number of packets successfully transmitted for two consecutive slots after type-1 collision. A first-order autoregressive model with weighting factor $\theta\in(0,1)$ is used for estimating $\lambda_t$; whenever a success by a single packet transmission in a slot, or type-1 collision occurs, one or $s$ packets are successfully transmitted.  If this is divided by the number of slots between two consecutive open slots, denoted by $L_{t-1}$,  it becomes the output rate from the system, i.e., throughput in the unit of packets per slot.  The estimation on $\lambda_t$ in line 2 is based on the assumption that the output rate would be equal to the input rate in equilibrium.

As shown in lines 3, 5, and 7, the BS observes the channel outcome of every open slot and then estimates the (average) number of backlogged users $\nu_t$ in lines 4, 6, and 8 according to the channel outcome. Particularly, $L_t $  in lines 6 and 8 can take three slots and one slot, depending on type-1 or type-2 collision, respectively.  In line 10, $\nu_t$ is updated by the estimated new arrivals for $L_t$ slots.  In line 11, the BS broadcasts the throughput-optimal (re)transmission probability $p_{t+1}^*$ for the next open slot so that the backlogged users apply it.  Therefore, instead of assuming a particular traffic arrival model, Algorithm \ref{algo1} estimates the number of backlogged users $\nu_t$ in an online manner and controls the transmission probability $p_t$ accordingly.  Consequently, Algorithm \ref{algo1} has capability of adapting to traffic arrival dynamics.

%the underlying idea is Bayesian optimation in the sense that 

We derive the update equations in lines 4, 6, and 8 in Algorithm \ref{algo1}. Hereafter, the subscript $t$ of $\nu_t$ is dropped for convenience. Let us remind that
the number of backlogged users is not known to the BS \emph{a prior}. The BS estimates the number of backlogged users by a Poisson distribution with mean $\nu$ as the \emph{a prior} distribution (or belief). Then, we can write the probability that the number of backlogged users is $k$ as
\begin{align}\label{eq:prior}
	\Phi_n(\nu)=\frac{\nu^n}{n!}e^{-\nu}.
\end{align}
In lines 4, 6 and 8, $\kappa=\nu p$ is a system parameter that depends on $K$, which is given in Remark \ref{Re:Re1} at the end of this subsection 
 
Based on the BS's belief that there are $n$ backlogged users, the joint probability that an idle slot occurs when $n$ users are backlogged is then expressed as
	\begin{align}\label{eq:eq29}
	\Pr[I,n]	=\mathcal{B}_{0}^{n}(p)\Phi _n(\nu).
\end{align}
The probability of a slot being idle is expressed as
\begin{equation}\label{eq:eq30}
	\Pr[ I ] =\sum_{n=0}^{\infty}\Pr[I,n] =e^{-\nu p}.
\end{equation} 
Once the slot is found idle, the BS can update its previous belief for the next open slot by using the \emph{a posteriori} probability of $n$ backlogged users given the idle slot:
%and the backlogged users have used the retransmission probability $p$ broadcast before:
\begin{align}\label{eq:eq35}
	\Pr[ n | I ] =\frac{\Pr\left[ I,n \right]}{\Pr\left[ I \right]}=\Phi _n\left( \nu \left( 1-p \right) \right).
\end{align}
Then, the expectation of \eqref{eq:eq35} is obtained as
\begin{align}\label{eq:eq34}
	\mathbb{E}[n|I]=\sum_{n=0}^{\infty}n\Pr[n|I]=\nu(1-p).
\end{align}
In line 4, we have used $\nu(1-p)=\nu-\kappa$.

By the same token, the joint probability of a successful slot by a single packet transmission with $n$ backlogged users can be expressed as
\begin{align}\label{eq:eq31}
	\Pr[ S,n]=\mathcal{B}_1^n(p)\Phi_n(\nu).
\end{align}
Similar to \eqref{eq:eq30}, we have the probability of a successful slot as 
\begin{equation}\label{eq:eq32}
	\Pr[ S] =\sum_{n=1}^{\infty}\Pr[ S,n] =\nu pe^{-\nu p}.
\end{equation}
Provided that the channel outcome is success, the BS's belief is updated by the a posteriori probability of $n$ backlogged:
\begin{align}\label{eq:eq37}
	\Pr[ n| S ] =\frac{\Pr[ S,n]}{\Pr[ S]}=\Phi _{n-1}\left( \nu \left( 1-p \right) \right).
\end{align}
The expectation of \eqref{eq:eq37} is obtained as
\begin{align}\label{eq:eq36}
	\mathbb{E}[n|S]=\nu(1-p)+1.
\end{align}
In this case, we subtract one from \eqref{eq:eq36} for the update equation in line 4, which accounts for the packet successfully transmitted. 
 \begin{figure}[pt]  \centering
	\subfigure[$\kappa$ vs. $K$.]{	
		\includegraphics[width=3.3in,height=2.6in]{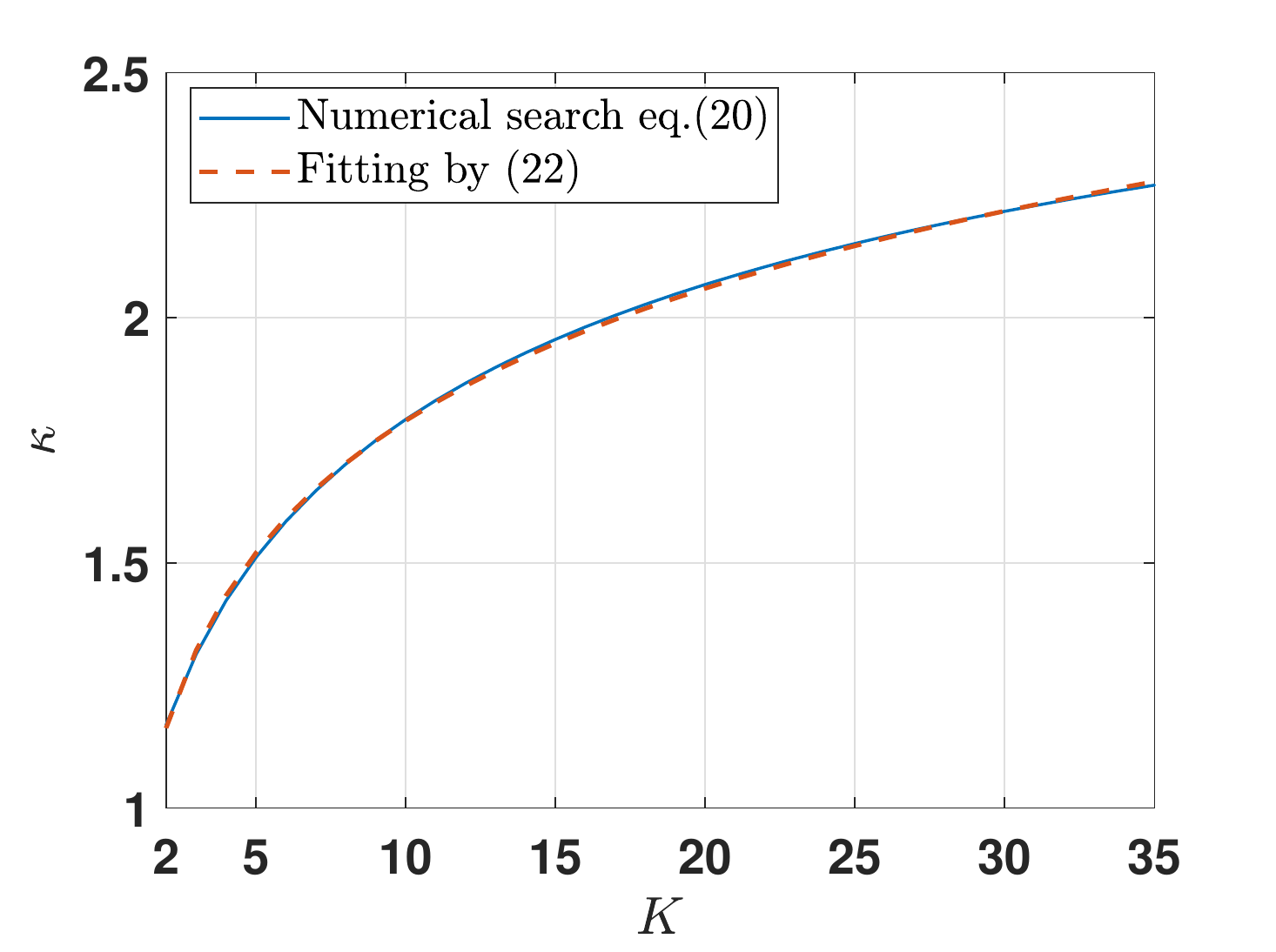}	\label{fig:fig22a}}\\
	\subfigure[$\tilde{\tau}^*$ vs. $K$.]{	
		\includegraphics[width=3.3in,height=2.6in]{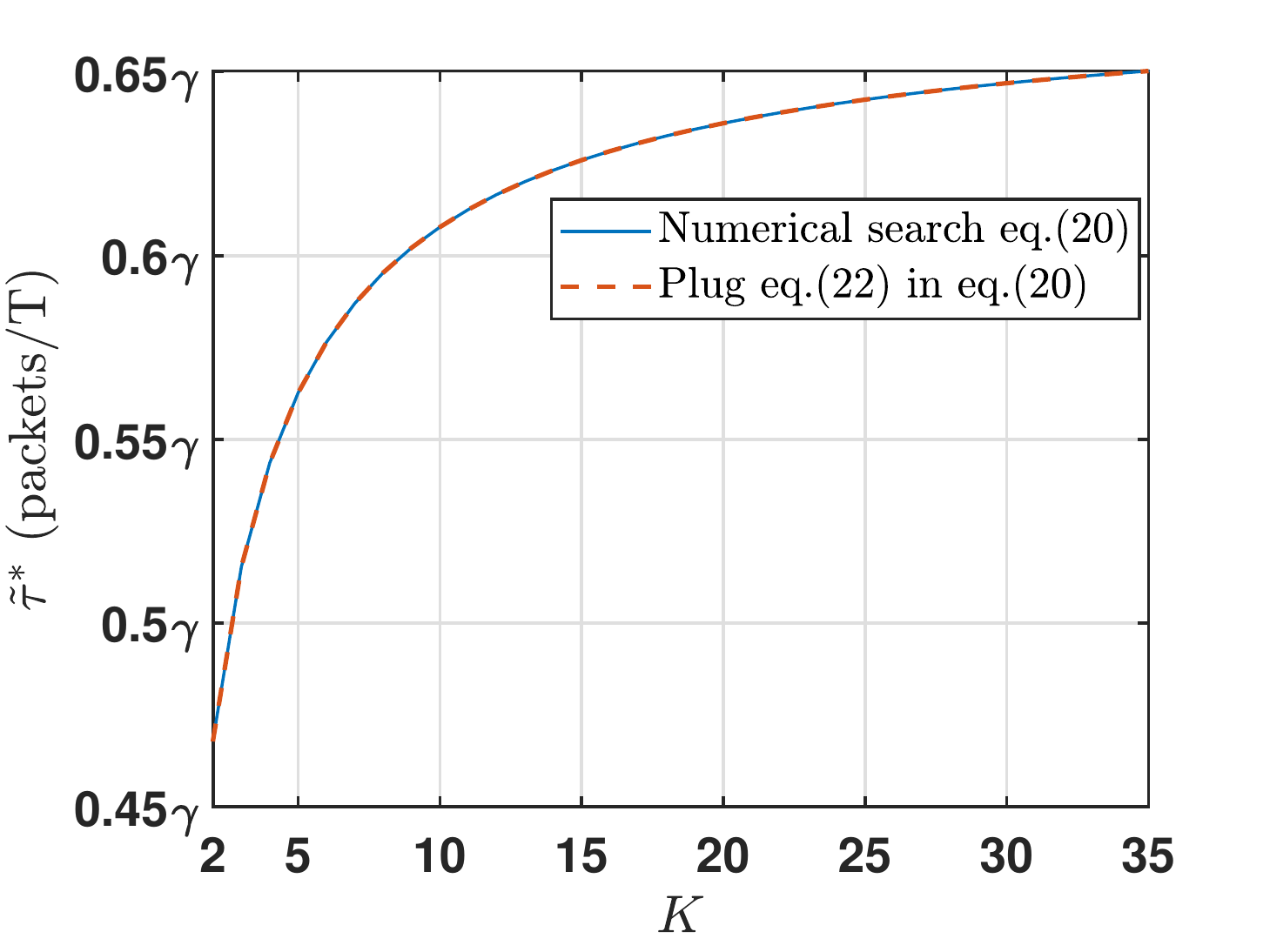}	\label{fig:fig22b}}
	\caption{Relationship between $\kappa$, $\tilde{\tau}^*$ and $K$.}		\label{fig:fig22}
\end{figure}  

%The essence of this algorithm is that the BS estimates the mean number of backlogged users, $\nu$ and broadcasts $p^*$, which is a function of $\nu$.  Notice that The proposed backoff algorithm is based on  By observing the channel event, e.g., idle, success or collision, it calculates the mean of the a posteriori distribution. 
%
%
%We assume infinite users with single buffer. Packet generation rate is $\lambda$.
%
%We suppose the prior distribution of the backlog size $n$ is a Poisson distribution with mean $\nu$.
%Probability of idle, success and collision in a cycle are same as ideal system.
%\begin{tabular}{|c|c|c|c|c|c|}
%	\hline
%	$K$&2&4&8&16&32\\
%	\hline
%	Optimal $\nu p$&1.1704&1.4233&1.7019&1.9814&2.2398\\
%	\hline
%	Max throughput&$\frac{0.4681}{T+\alpha}$&$\frac{0.5436}{T+\alpha}$&$\frac{0.5953}{T+\alpha}$&$\frac{0.6284}{T+\alpha}$&$\frac{0.6484}{T+\alpha}$\\
%	\hline
%\end{tabular}
%
%Posterior distribution of $\nu$ is same as the ideal system.
%Mean of the posterior distribution of $\nu$ is also same as the ideal system.
%
%where $s$ is number of successes in a collision cycle.
%
%We assume infinite users with single buffer. Packet generation rate is $\lambda$.
%
%We suppose the prior distribution of the backlog size $n$ is a Poisson distribution with mean $\nu$.
%Probability of idle, success and collision in a cycle are:
%$for n saturated user, \nu p=\frac{n}{\frac{n}{0.2702*log_{2}K+0.8937}-0.01257(log_{2}K)^{2}+0.21log_{2}K-0.2899}$

It is noteworthy that the a posteriori distributions  in \eqref{eq:eq35} and \eqref{eq:eq37} are also a Poisson distribution with mean rate $\nu(1-p)$, the same form of the a prior distribution.  

Let us recall that $\mathbb{C}$ indicates a collision regardless of  type-1 or  type-2. The joint probability that (re)transmissions from $n$ backlogged users result in a collision is expressed as 
\begin{align}
	\Pr[\mathbb{C},n]=(1-\mathcal{B}_0^n(p)-\mathcal{B}_1^n(p))\Phi_n(\nu).
\end{align}
The probability that a collision happens in a slot is expressed as
\begin{equation}
	\Pr[ \mathbb{C}] =\sum_{n=1}^{\infty}\Pr[\mathbb{C},n] =1-(1+\nu p)e^{-\nu p}.
\end{equation} 
The a posteriori probability of $n$ backlogged users found given a collision slot is expressed as 
\begin{align}\label{eq:eq41}
 &	\Pr[ n| \mathbb{C} ]=\frac{\Pr[ \mathbb{C},n]}{\Pr[ \mathbb{C}]}=\frac{(1-\mathcal{B}_0^n(p)-\mathcal{B}_1^n(p))\Phi_n(\nu)}{1-(1+\nu p)e^{-\nu p}}\\&~~~=\frac{\Phi_n(\nu)-e^{-\nu p}\left(\Phi_{n}(\nu(1-p))+\nu p\Phi_{n-1}(\nu(1-p))\right)}{1-(1+\nu p)e^{-\nu p}}.\notag
\end{align}
Although \eqref{eq:eq41} is no longer a Poisson distribution, the BS still holds the belief that it would be a Poisson distribution with mean
\begin{align} \label{eq:eq_collision}
	\mathbb{E}[n|\mathbb{C}]=\nu + \frac{(\nu p)^2}{e^{\nu p}-\nu p - 1}.
\end{align}
\begin{table}[]\centering
\renewcommand{\arraystretch}{1.3}
	\caption{The Optimal $\kappa$ with Various $K$'s}
	\begin{tabular}{c!{\vrule width 0.6 pt}  l l l l l}
		\Xhline{0.8 pt}
		$K$             & \multicolumn{1}{c}{2}      & \multicolumn{1}{c}{4} & \multicolumn{1}{c}{8} & \multicolumn{1}{c}{16} & \multicolumn{1}{c}{32} \\ \Xhline{0.6 pt}
		$\kappa=\nu p$ & 1.1704                      & 1.4233                 & 1.7019                 & 1.1914                  & 2.2398                  \\ \Xhline{0.2 pt}
		$\tilde{\tau}^*$ & {$0.4681 \gamma$} &              {$0.5436 \gamma$}         &       {$0.5953 \gamma$}                &      {$0.6284\gamma$}            &      {$0.6484 \gamma$}                 \\ \Xhline{0.8 pt}
	\end{tabular}\label{Tab1}
\end{table}\begin{figure*}[pt]  \centering
\subfigure[Maximum throughput with $K$.]{	
	\includegraphics[width=3.3in,height=2.65in]{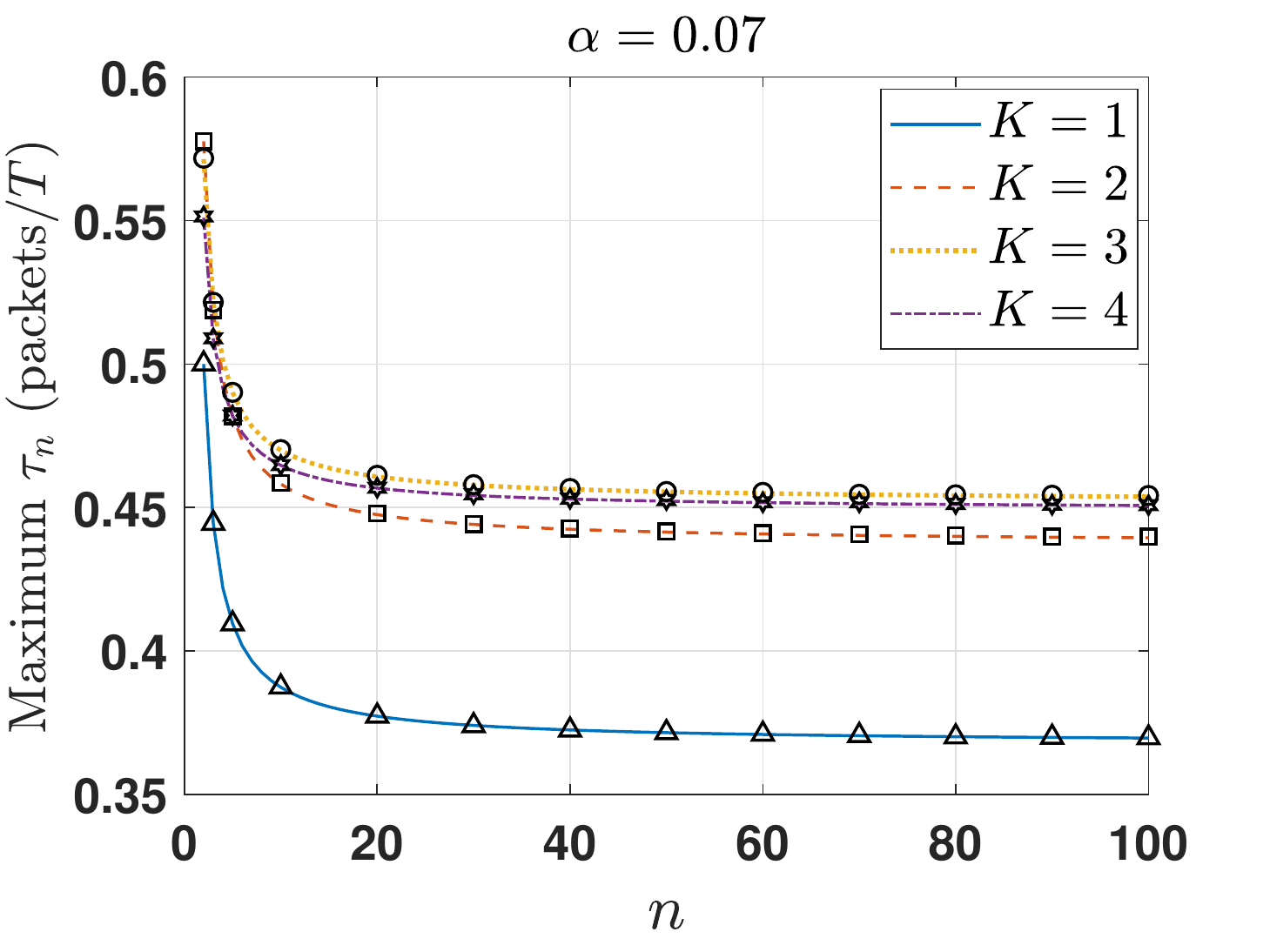}	\label{fig:fign1}}	\subfigure[Maximum throughput with $K$.]{	
	\includegraphics[width=3.3in,height=2.65in]{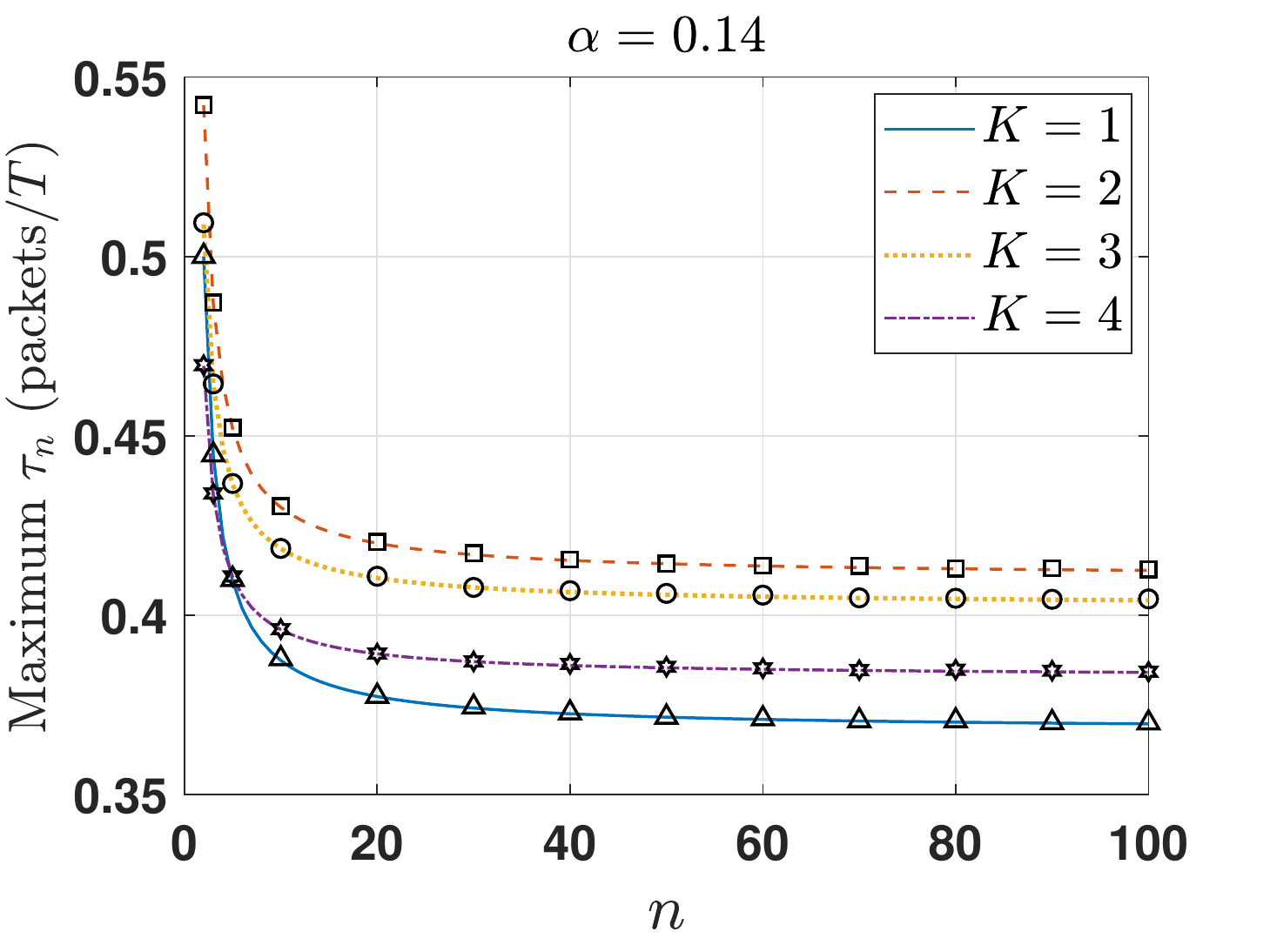}	\label{fig:fign4}}\\
\subfigure[Throughput with small population sizes.]{	
	\includegraphics[width=3.3in,height=2.65in]{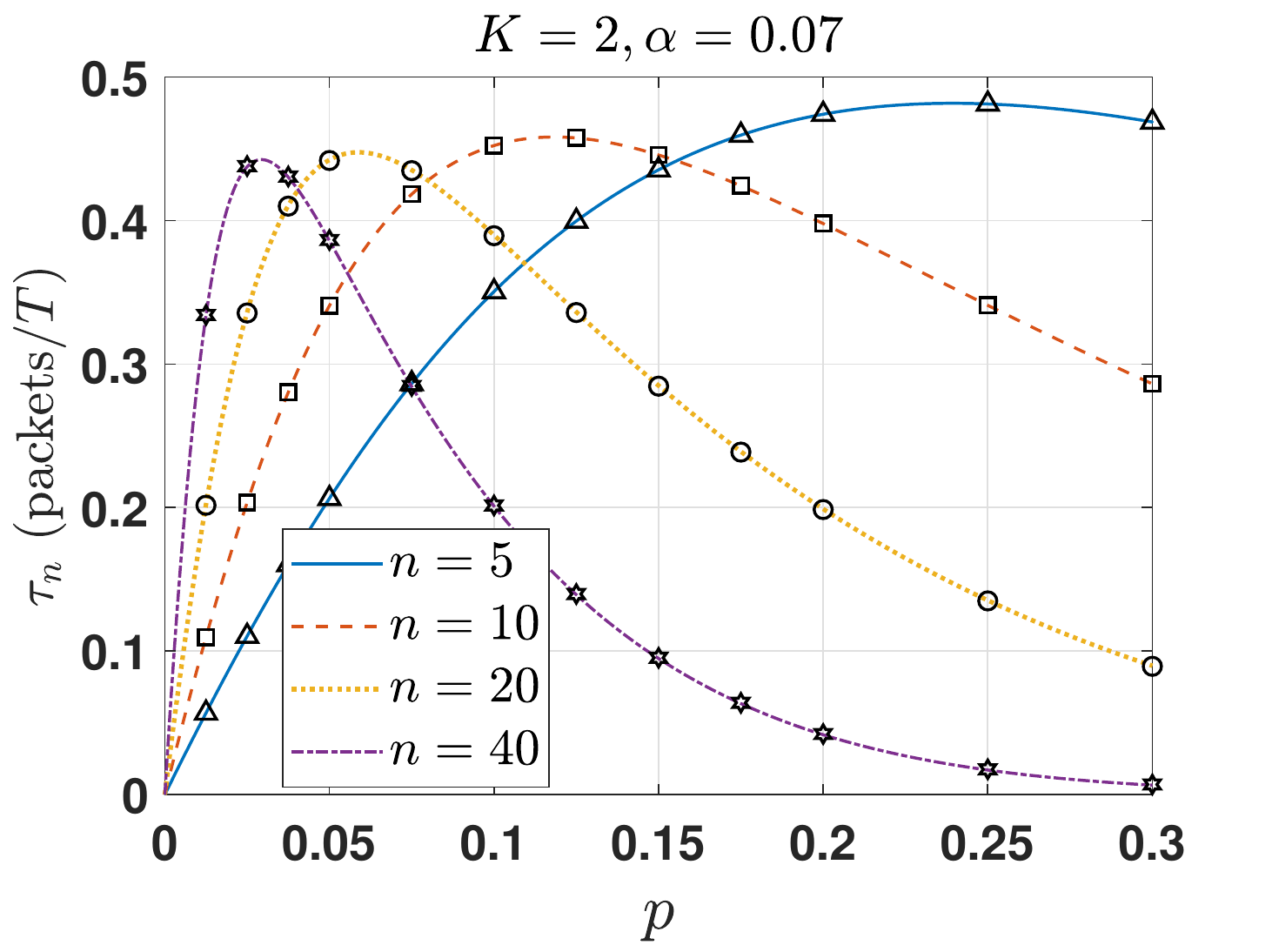}	\label{fig:fign2}}\subfigure[Throughput with small population sizes.]{	
	\includegraphics[width=3.3in,height=2.65in]{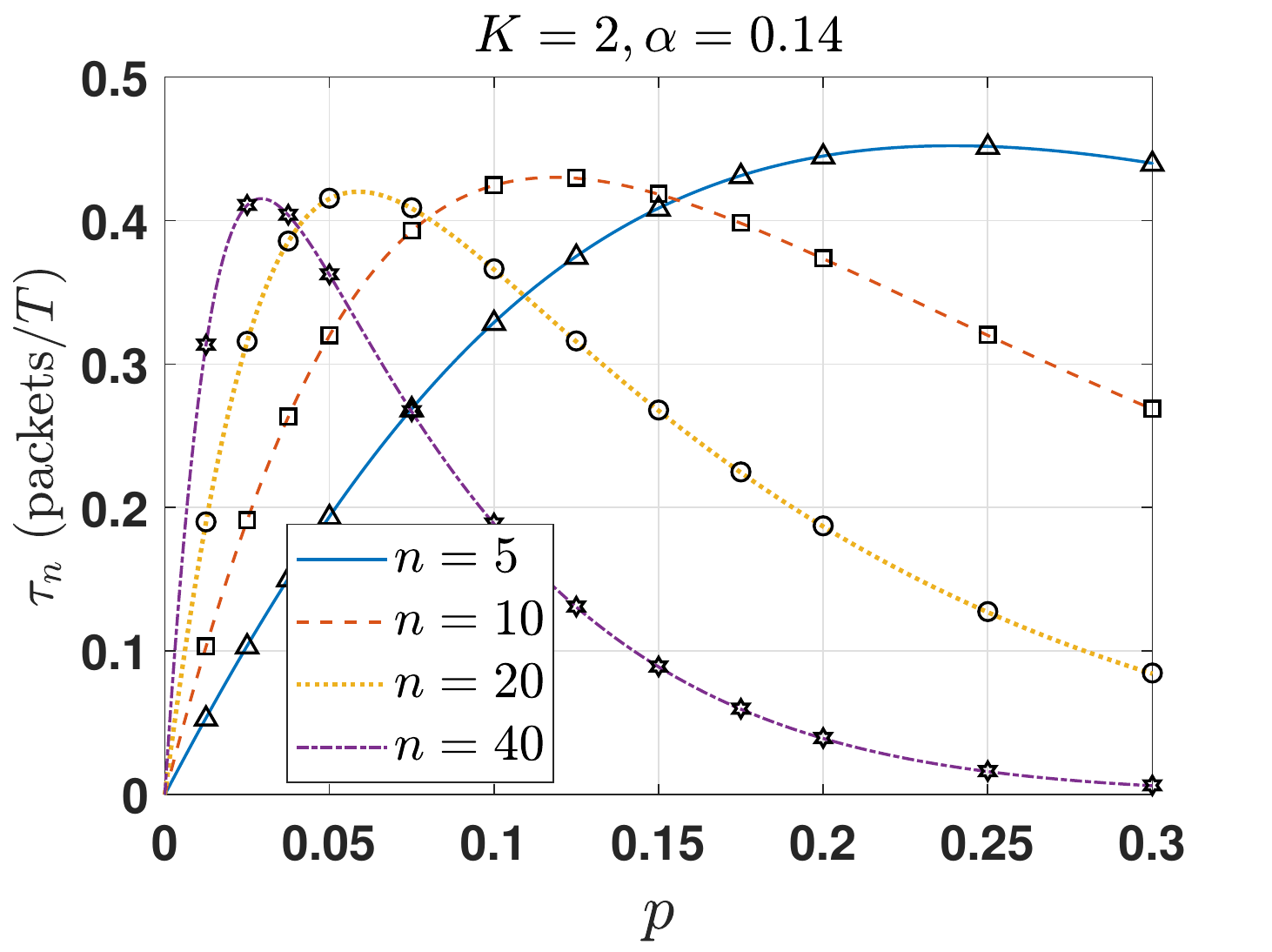}	\label{fig:fign5}}\\
\subfigure[Throughput with $n=100$]{
	\includegraphics[width=3.3in,height=2.65in]{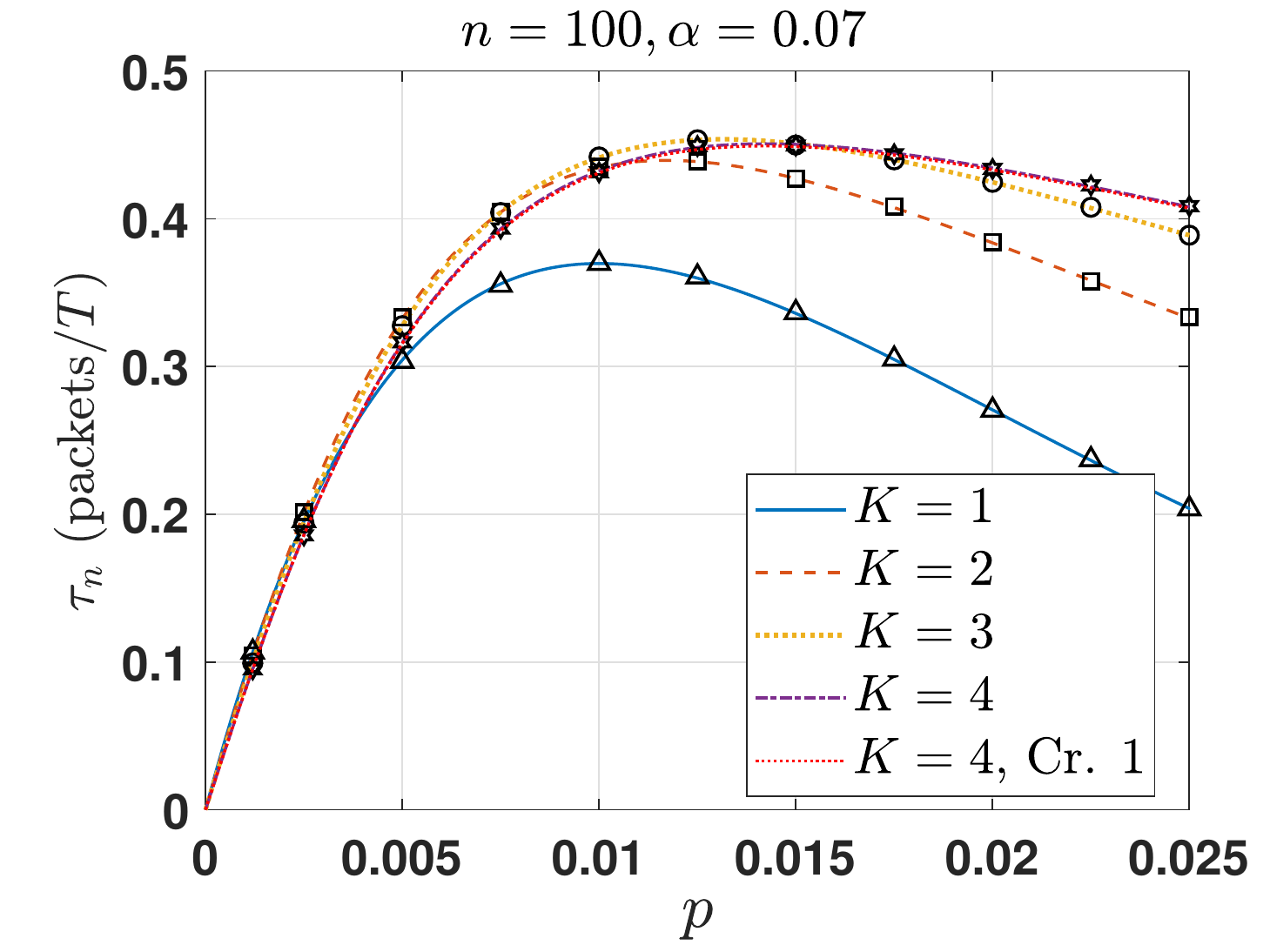}	\label{fig:fign3}}	\subfigure[Throughput with $n=100$]{
	\includegraphics[width=3.3in,height=2.65in]{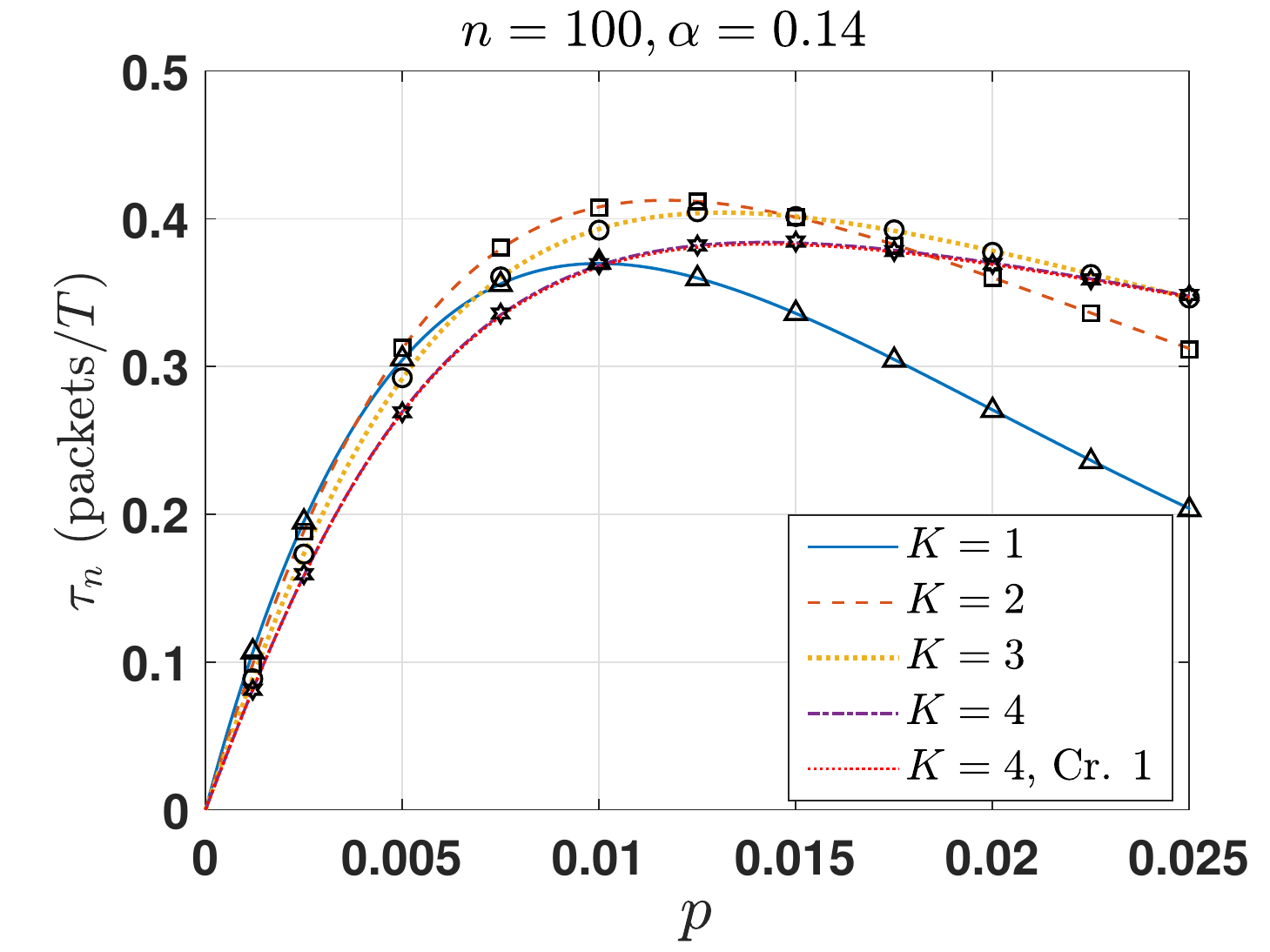}	\label{fig:fign6}}
%		\subfigure[Throughput with large population sizes..]{
	%		\includegraphics[width=2.5in,height=1.9in]{fig1}	\label{fig:fign4}}
%		\subfigure[The average idle period.]{	
	%		\includegraphics[width=2.5in,height=1.9in]{fig61}	\label{fig:fig7}}
%\vspace{-0.3cm}
%	\subfigure[Throughput of unslotted ALOHA system.]{	\includegraphics[width=2.5in,height=1.9in]{fig4}	
	%		\label{fig:fig3}}
\caption{Throughput behavior of S-ALOHA system with $\alpha=0.07$ and $0.14$.}		\label{fig:fig3}
%	\caption{Throughput with $T=1$ and various (small) population sizes.}
\end{figure*}
To find $\kappa$ in lines 4, 6 and 8, let us construct the following lemma.
\begin{Lemma}\label{Lm:Lm1}
Suppose that the BS estimates the number of backlogged users by a Poisson distribution with mean rate $\nu$. Throughput under this assumption, denoted by $\tilde{\tau}_{\nu}$ (packets/$T$), is expressed as
\begin{align} \label{eq:eq20a}
	\tilde{\tau}_\nu
=\gamma\frac{\nu p e^{-\nu p }+\frac{2 \nu p}{K}\Bigg[\frac{e^{-\frac{\nu p}{K}}-e^{-\nu p}}{1-e^{-\frac{\nu p}{K}}}-(K-1) e^{-\nu p}\Bigg]}{3-2 e^{-\nu p }\left(K \left(e^{\frac{\nu p}{K}}-1\right)+1\right)},
\end{align}
where $\gamma = T / T_s$.
\begin{IEEEproof} Using renewal reward theorem, we can write $\tilde{\tau}_\nu$ as 
	\begin{align}\label{eq:eq28}
\tilde{\tau}_\nu = \frac{\sum_{n=1}^\infty\mathbb{E}[R|n]  \Phi_{n}(\nu)}{ \sum_{n=0}^\infty\mathbb{E}[Z|n]  \Phi_{n}(\nu) / T} .
	\end{align}
Due to the Poisson distribution of the backlogged users, both the numerator and the denominator are averaged over the number of backlogged users.  The normalization by $T$ appears in the denominator because the throughput has the unit of packets per $T$. 
%	in which we have used \eqref{eq:eq6aa}, \eqref{eq:eq11}, and \eqref{eq:eq19}. \\	
%	We need to find $	\Pr[ S_{1,\varphi}] $ and $\Pr[C_2]$ in \eqref{eq:eq28}. Using  \eqref{eq:eq9} in $n$-user system we have
%	\begin{align}\label{eq:eq33}
%		\Pr[ S_{1,\varphi}] =\sum_{n=2}^{\infty}{\sum_{i=2}^n{\sum_{j=1}^{K-1} \frac{i\mathcal{B}_{i}^{n}(p)}{K^i}}}\left( K-j \right) ^{i-1}\Phi_n( \nu ).
%	\end{align}
%	From \eqref{eq:eq18}, the probability of type-2 collision is expressed as
%	\begin{align}\label{eq:eq34m}
%		\Pr[ C_2 ] =\sum_{n=2}^{\infty}{\sum_{i=2}^n \frac{\mathcal{B}_{i}^{n}(p)}{K^{i-1}}}\Phi _n\left( \nu \right).
%	\end{align}
	Putting  \eqref{eq:eq3} and  \eqref{eq:eq_renewal_length} into \eqref{eq:eq28} yields the result.	 Interestingly, the expression of $\tilde{\tau}_\nu$ becomes $\tau_\infty$ in \eqref{eq:eq22} if we set $\eta = \nu p$.  
	\end{IEEEproof}
\end{Lemma}

  Let $\kappa=\nu p$ and it is a maximizer of $\tilde{\tau}_\nu$ in Lemma \ref{Lm:Lm1}. Table \ref{Tab1} lists $\kappa$ values according to $K$, whereas $\tilde{\tau}^*$ indicates $\tilde{\tau}_\nu$ with $\kappa$. We obtain Table \ref{Tab1} by numerical search for \eqref{eq:eq20a}. To see a relationship between $\kappa$, $\tilde{\tau}^*$ and $K$, the following remark can be given:
\begin{Remark}\label{Re:Re1}
	For $2\leq K\leq 32$, the throughput-optimal retransmission is numerically obtained as $p_\nu^*=\frac{\kappa}{\nu}$, where constant $\kappa$ is expressed as
\begin{align}		\label{eq:eq22a}	\kappa 
			=	0.2697\log_{2}K+0.8943.
			\end{align}
	%		\begin{cases}
	%			1,& \text{if } K=1\\
	%			0.2697\log_{2}K+0.8943,              &  \text{else if}\;\; 2\leq K\leq 32
	%		\end{cases}
Plugging $\kappa$ into $\tilde{\tau}_\nu$ yields 
%\begin{align}
%	\tilde{\tau}^*=\gamma(0.1075\log_2K-0.01038(\log_2K)^2+0.368).\end{align}
%	\textcolor{red}{
	\begin{align}
	\tilde{\tau}^*=\gamma\left(0.673-\frac{0.7828}{K+1.904}\right).\end{align}%}
Especially for $K=1$, we have $\kappa=1$.  In Fig. \ref{fig:fig22}, it is interesting to see that the throughput increases  in $K$ when optimal $\kappa$ is used, and converges to $0.673\gamma$. Note that the optimal parameter $\kappa$ we use in Section \ref{sec:num} is found numerically from \eqref{eq:eq20a} instead of \eqref{eq:eq22a}.
\end{Remark}

The update equations for $\nu_t$ in lines 4, 6, and 8 are obtained by plugging $\kappa$ into \eqref{eq:eq34}, \eqref{eq:eq36}, and \eqref{eq:eq_collision}, respectively. It is necessary to know how large the packet arrival rate $\lambda$ (packets/$T$) can be accepted by the proposed Bayesian backoff algorithm so long as the number of backlogged users grows infinite. 
\begin{Lemma} \label{Lm:lm2} S-ALOHA with TO is stabilized by Bayesian backoff algorithm since 
\begin{align}
	\lambda &< \min_{m\geq 1} \;\tau_m(p_m^*),
\end{align}
where $p_m^*$ is given in Remark 1.
\begin{IEEEproof}
See Appendix \ref{App4}.
	\end{IEEEproof}
\end{Lemma}

\begin{Corollary}\label{Cr:Cr4}
S-ALOHA with TOs is unstable when the users make use of a fixed retransmission probability $p$.  \begin{IEEEproof} See Appendix \ref{App5}.
	\end{IEEEproof}
\end{Corollary}
Corollary \ref{Cr:Cr4} shows importance of Bayesian backoff algorithm to stabilize the system.

 \begin{figure}[pt]  \centering
\subfigure[Throughput vs. $\alpha$.]{	
	\includegraphics[width=3.3in,height=2.6in]{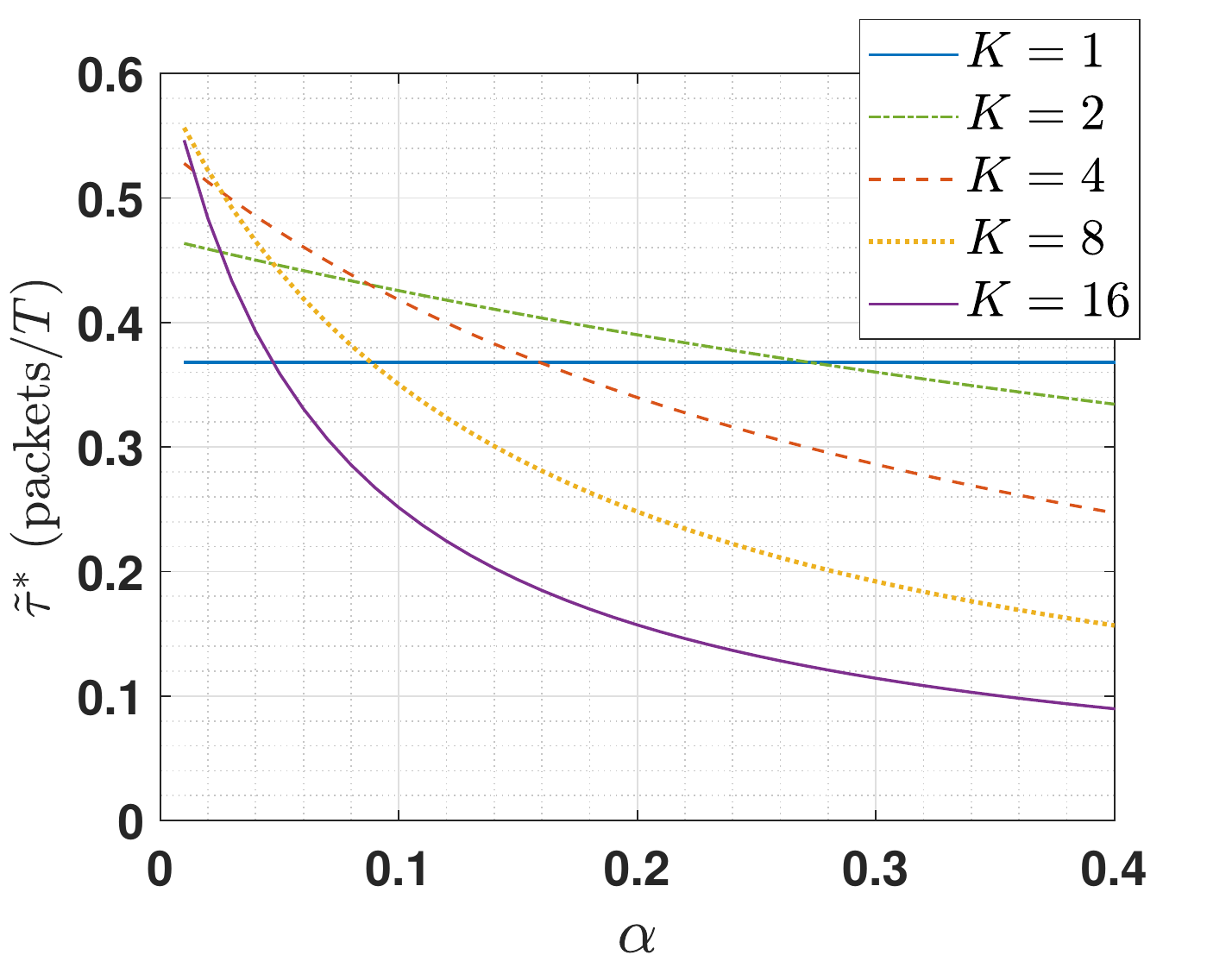}	\label{fig:fign16}}\\
\subfigure[Throughput vs. $K$.]{	
	\includegraphics[width=3.3in,height=2.6in]{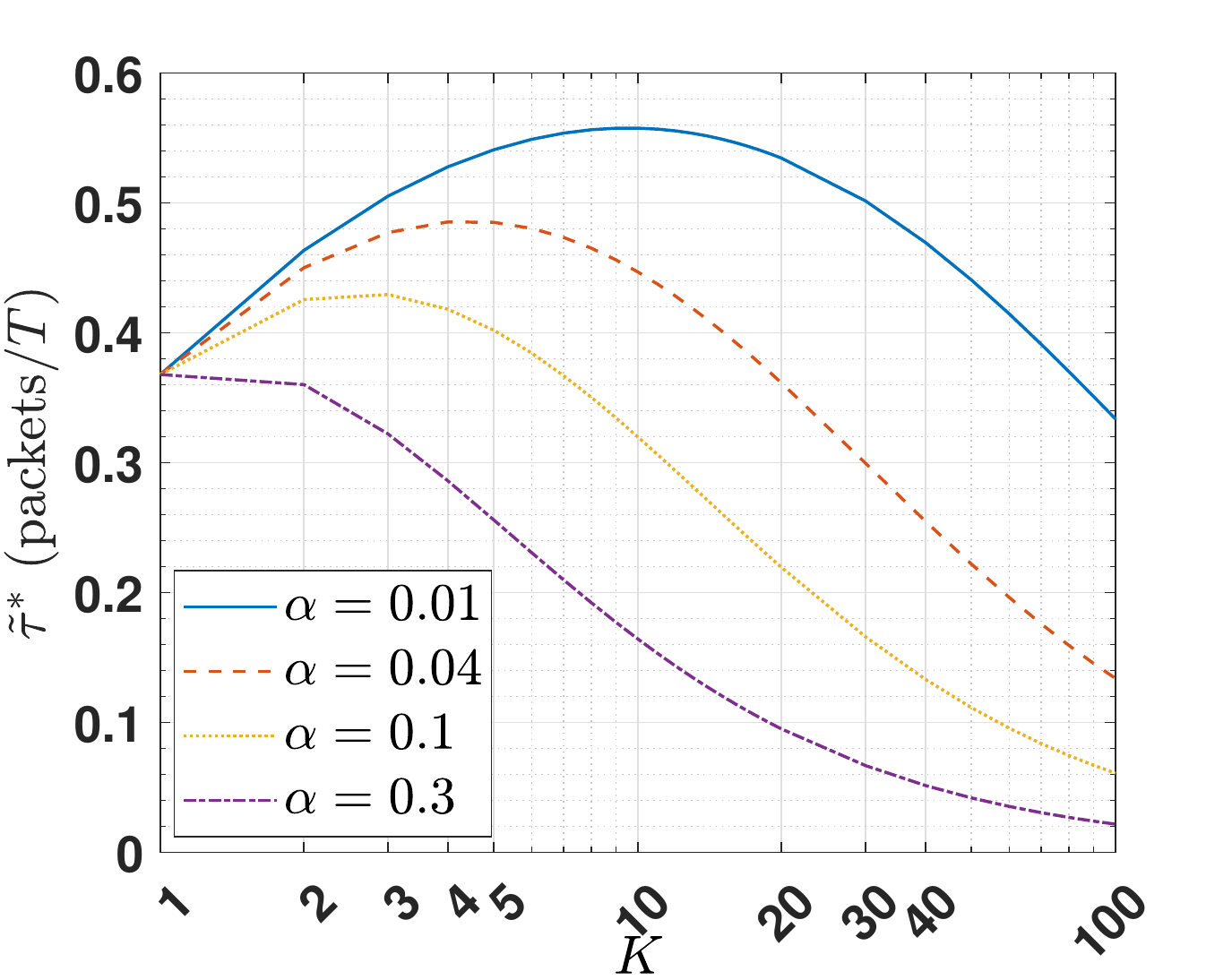}	\label{fig:fign17}}
\caption{Throughput behavior with various $\alpha$'s and $K$'s.}		\label{fig:fig9}
\end{figure}  
\subsection{Algorithm with Uniform Backoff Window}
Until now, we have assumed that the BS broadcasts retransmission probability $p_t^*$ at each open slot and the backlogged users apply it for retransmission. As mentioned before, if not retransmitting with probability $1-p_t^*$, they need to look for another retransmission at the next open slot. It can lead to more energy consumption, since the backlogged users keep on watching the broadcast message until transmitting.  We consider the Bayesian backoff algorithm with uniform window. In this system, instead of $p_{t+1}^*$ in line 11, the BS broadcasts window size $U_{t+1}$. On receiving $U_{t+1}$, the backlogged users take on a counter randomly chosen between $0$ and $U_{t+1}-1$. If it is zero, those users retransmit if the slot is open. Otherwise,  the backlogged user counts it down by one every slot \emph{without monitoring} the broadcast message. 
With the count hitting zero, the backlogged user transmits if the slot is open. If not, i.e., closed slot, he does not retransmit and waits for the broadcast message at the end of the closed slot and repeats this with $U_{t+1}$ newly broadcast.

To determine $U_{t+1}$, the average retransmission interval of the algorithm with $p^*_{t+1}$ is $1/p_{t+1}^*$.  If the backlogged users use (uniform) window $U_{t+1}$, its average retransmission interval is $\frac{U_{t+1}}{2}$. By matching these averages, i.e., $\frac{U_{t+1}}{2}=1/p_{t+1}^*$ and making it an integer, we can find $U_{t+1}$ as\begin{align}\label{eq:eq55}
	U_{t+1}=\left\lceil\frac{2}{p_{t+1}^*}\right\rceil.
\end{align}   
%Using \eqref{eq:eq55}, we make the average interval of retransmissions of two algorithms equal. 
The advantage of this uniform window algorithm is that the BS broadcasts an integer-valued window size, instead of a real-valued probability. This is relatively easy to implement a few bits in a broadcast message. Moreover, the backlogged users do not need to monitor continuously the broadcast message.  %If a user has a packet to send and finds the slot idle, 

%\begin{figure}[pt]  \centering
%	\subfigure[Maximum throughput with $K$.]{	
%		\includegraphics[width=2.75in,height=2.2in]{max_throughput2_m2}	\label{fig:fign4}}\\
%	\subfigure[Throughput with small population sizes.]{	
%		\includegraphics[width=2.75in,height=2.2in]{throughput_p_small_pop2_m2}	\label{fig:fign5}}\\
%	\subfigure[Throughput with $n=100$]{
%		\includegraphics[width=2.75in,height=2.2in]{throughput_with_n_1002m_m2}	\label{fig:fign6}}
%	%		\subfigure[Throughput with large population sizes..]{
%	%		\includegraphics[width=2.5in,height=1.9in]{fig1}	\label{fig:fign4}}
%	%		\subfigure[The average idle period.]{	
%	%		\includegraphics[width=2.5in,height=1.9in]{fig61}	\label{fig:fig7}}
%	%\vspace{-0.3cm}
%	%	\subfigure[Throughput of unslotted ALOHA system.]{	\includegraphics[width=2.5in,height=1.9in]{fig4}	
%	%		\label{fig:fig3}}
%	\caption{Throughput behavior of S-ALOHA system with $\alpha=0.14$.}		\label{fig:fig4}
%	%	\caption{Throughput with $T=1$ and various (small) population sizes.}
%\end{figure}

\section{Numerical Results}\label{sec:num}
We build simulation with Matlab: Run time of each simulation is set to $10^6$ slots and time-averaged result is obtained. Unless otherwise specified, in the following figures, the lines and symbols indicate analysis and simulation results, respectively.  Throughout this section, we set the packet transmission time to one, i.e., $T=1$ for convenience. It is also notable that the system with $K=1$ is S-ALOHA without a TO.
%%%%%%%%%%%%%%%%%%%%%%%%%%%%%%%%%%%%%%%%%%%%%%%%%%%%%%%%%%%%%%%%%%%%%%%%%%%%%%%%%%%%%%%%%%%%%%%%%%%%%%%%%%%%%%%%%%%%%%%%%%%
%Recall that the system is reduced to (original) S-ALOHA system without TOs for $K=1$ and that since one slot length $T_s=(K-1)\alpha+T$, either increasing $K$ or $\alpha$ prolongs the slot length. We examine throughput for $\alpha=0.01$ in Fig. \ref{fig:fig3} and $\alpha=0.04$ in Fig. \ref{fig:fig4}, respectively, to see the effect of $\alpha$ and $K$. %It can be seen that analysis and simulation results show good agreements. 
%%%%%%%%%%%%%%%%%%%%%%%%%%%%%%%%%%%%%%%%%%%%%%%%%%%%%%%%%%%%%%%%%%%%%%%%%%%%%%%%%%%%%%%%%%%%%%%%%%%%%%%%%%%%%%%%%%%%%%%%%%%
\begin{figure}[pt]  \centering
	%	\subfigure[Throughput vs. $\alpha$.]{	
		\includegraphics[width=3.3in,height=2.6in]{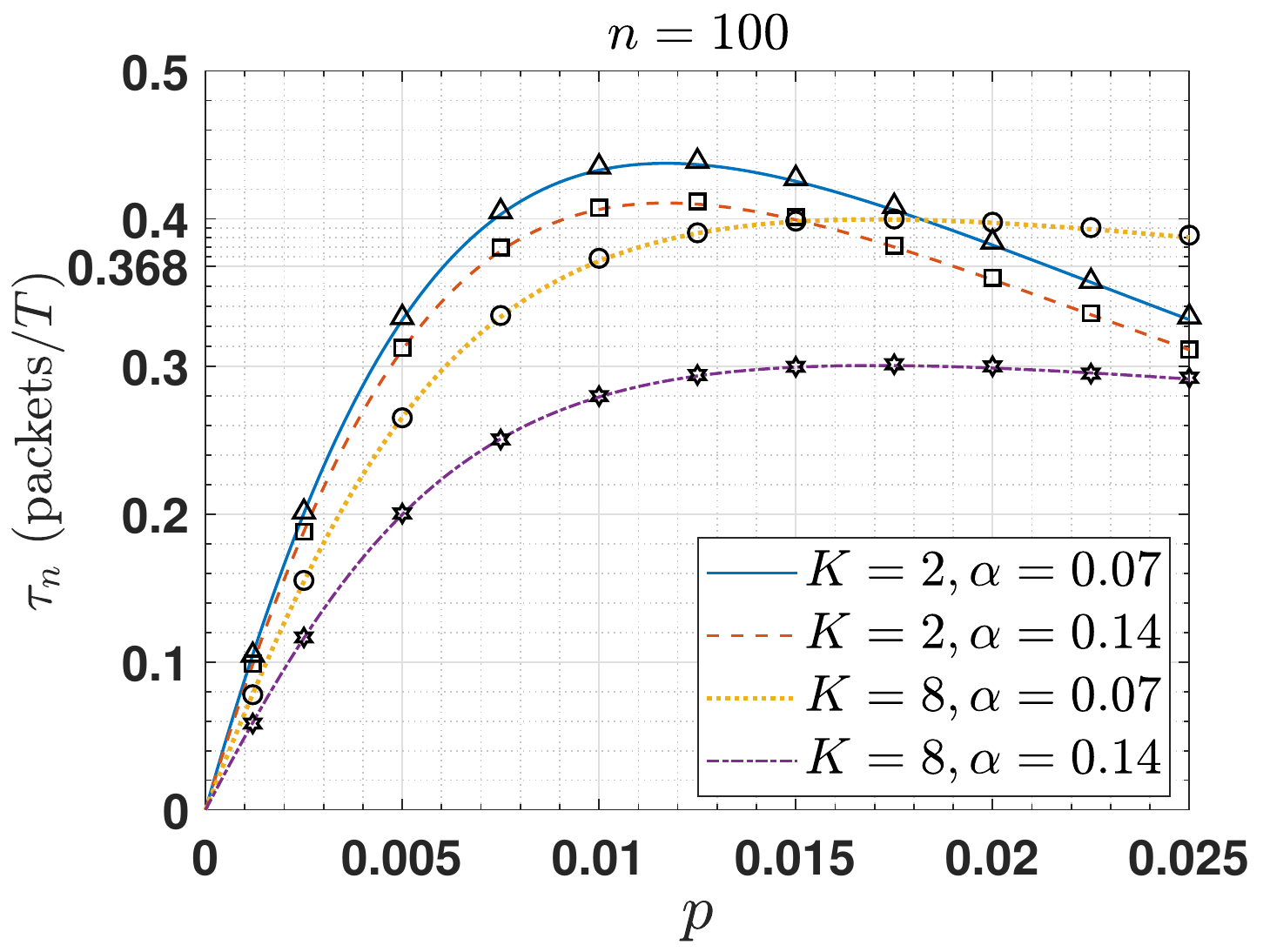}	%}
	%\\
	%\subfigure[Throughput vs. $K$.]{	
		%	\includegraphics[width=2.5in,height=1.92in]{max_throughput_K_alpha}	\label{fig:fign17}}
	\caption{Throughput approximation by Corollary 1.}		\label{fig:fig5_new}
\end{figure}

Let us examine the maximum throughput and the existence of optimal transmission probability by varying population size first. Figs. \ref{fig:fign1} and \ref{fig:fign4} show the maximum (achievable) throughput vs.  population size $n$ as $K$ increases for $\alpha = 0.07$ and $0.14$, respectively\footnote{As mentioned Section \ref{sys}, if one packet transmission takes one slot of 14 OFDM symbols and a TO has the length of one or two OFDMA symbols,  we have $\alpha=1/14\approx 0.07$ and $2/14\approx0.14$. It should be noted that the length of $\alpha$ is a relative value with respect to the packet transmission time.}.  As population size $n$ grows large, the (original) S-ALOHA system with $K=1$ shows $0.3679$-throughput. 
However, for $K=2$ and $3$, we can see higher throughput $\tau_n=0.441$ and $0.455$, respectively in Fig. \ref{fig:fign1}. Notice  that throughput is significantly high if very small population size $n\leq 10$. As $n$ exceeds 20, throughput converges to a constant value. In Figs. \ref{fig:fign2} and \ref{fig:fign5}, as in Remark \ref{Re:Re1}, the existence of throughput-maximizing (optimal) transmission probability $p$ can be expected for each $n$. For example, the throughput with $n=40$ can be maximized at $p=0.03$  in Fig. \ref{fig:fign3} and $p=0.0293$ in Fig. \ref{fig:fign6}, respectively. It is notable that as $n$ increases, the throughput becomes sensitive to $p$.   Figs. \ref{fig:fign3} and \ref{fig:fign6} illustrate how much throughput can increase for $n=100$ as $K$ increases. In both figures, $K=3$ improves the throughput as good as $K=4$, because $K=3$ brings sufficient contention resolution effect for $n=100$ with reasonable efficiency. 

\begin{table*}[pt]\centering
	\caption{Optimal System Parameters with Various $\alpha$'s}
	\renewcommand{\arraystretch}{1.3}
	\begin{tabular}{c!{\vrule width 0.6 pt} c c c c c c c}
		\Xhline{0.8 pt}
		$\alpha$   & \multicolumn{1}{c}{0.01}  & \multicolumn{1}{c}{0.02}  & \multicolumn{1}{c}{0.03}  & \multicolumn{1}{c}{0.04}  & \multicolumn{1}{c}{0.07}  & \multicolumn{1}{c}{0.14}  & \multicolumn{1}{c}{0.21} \\ \Xhline{0.6 pt} 
		Optimal $K$   & 10   & 7     & 5     & 4     & 3     & 2     & 2  \\ \Xhline{0.2 pt}
		$\kappa=\nu p$   & 1.7927   & 1.6474    & 1.5115   & 1.4233   & 1.3136   & 1.1704   & 1.1704  \\ \Xhline{0.2 pt}
		$\tilde{\tau}^*$   & 0.5576   & 0.5241    & 0.5024   & 0.4854   & 0.4521   & 0.4107   & 0.3869   \\ \Xhline{0.8 pt}
	\end{tabular}\label{Tab2}
\end{table*} 
	Now, let us consider how $K$ and $\alpha$ affect the maximum throughput. In Fig. \ref{fig:fign1}, when $K$ is increased up to $4$ from $3$, throughput goes down slightly. In Fig. \ref{fig:fign4} it can be seen that as $\alpha$ increases twice, the  $K$ of yielding the highest throughput reduces from 4 to 2.  This is because increasing $K$ brings about contention resolution effect, but too many $K$'s can lower transmission efficiency.  %The throughput is examined for various small population sizes with $K=2$ in Figs. \ref{fig:fign2} and \ref{fig:fign5}. On the other hand, 
	 In this sense, 
In Fig. \ref{fig:fig3}, it can be seen that the smaller the $\alpha$, the higher throughput. For $\alpha=0.07$, throughput goes up to $0.45$, while it is below $0.45$ for $\alpha=0.14$. In Figs. \ref{fig:fign4} although throughput with $K=4$ is better than S-ALOHA system without TOs, it is much lower than that with smaller $K$'s. Since a large $\alpha$ can lower transmission efficiency, keeping increasing $K$ does not necessarily bring throughput enhancement.

%For two or four TOs, the maximum throughput can read $0.4673$, and $0.531$, respectively. While $15$ percent of a slot is used for TOs for $K=8$, throughput is remarkably enhanced up to $0.5577$. 
 %From Fig. \ref{fig:fign3} we can infer that  if $p$ is controlled as a function of $n$ over time, the system can achieve the maximum throughput, which is the objective of Bayesian backoff algorithm. 

% \textcolor{blue}{From Fig. \ref{fig:fig3} to \ref{fig:fig4},  The overall throughput behaviors observed in Fig. \ref{fig:fig3} can be also found in Fig. \ref{fig:fig4} except for slightly reduced maximum throughput.}
%\begin{figure*}[pt]

	\begin{figure}[pt]  \centering
	\vspace{-0.1cm}
		\subfigure[Average access delay vs.  arrival rate $(\alpha = 0.07)$.]{	
			\includegraphics[width=3.3in,height=2.65in]{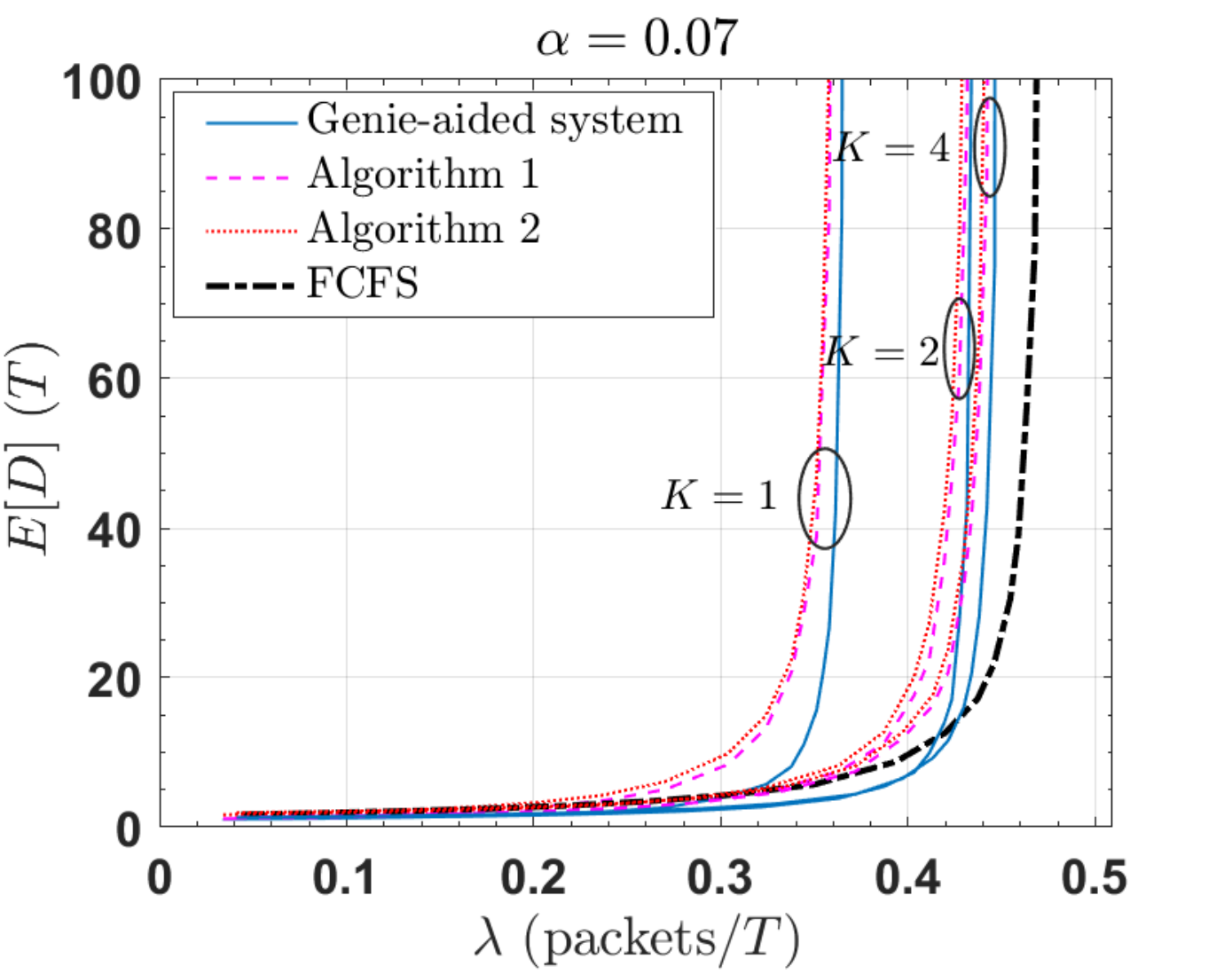}	\label{fig:fign8m}}\\
		\subfigure[Average access delay vs.  arrival rate $(\alpha = 0.04)$.]{	
			\includegraphics[width=3.3in,height=2.65in]{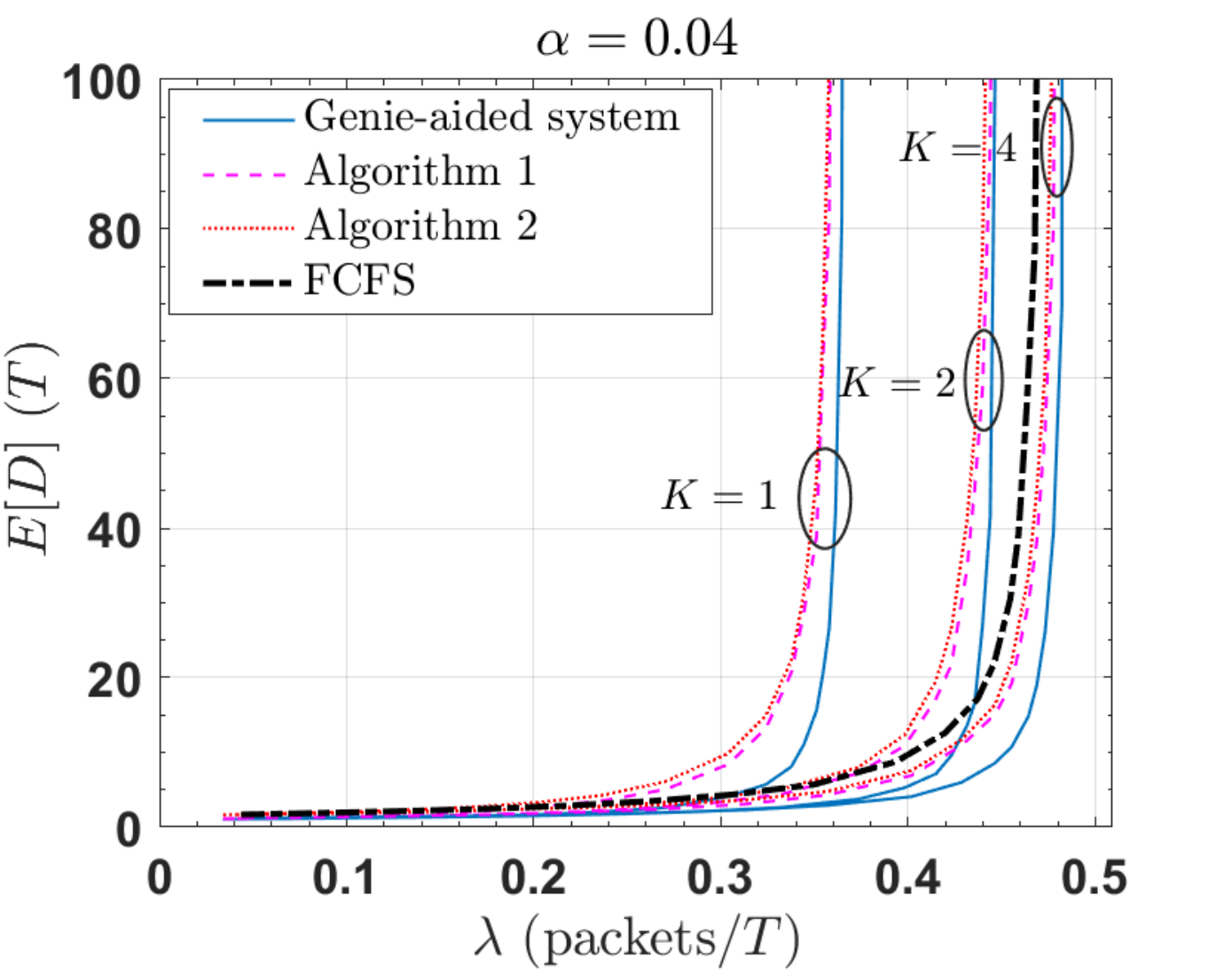}	\label{fig:fign8}}\\
		\subfigure[Average access delay vs.  arrival rate $(\alpha = 0.01)$.]{	
			\includegraphics[width=3.3in,height=2.65in]{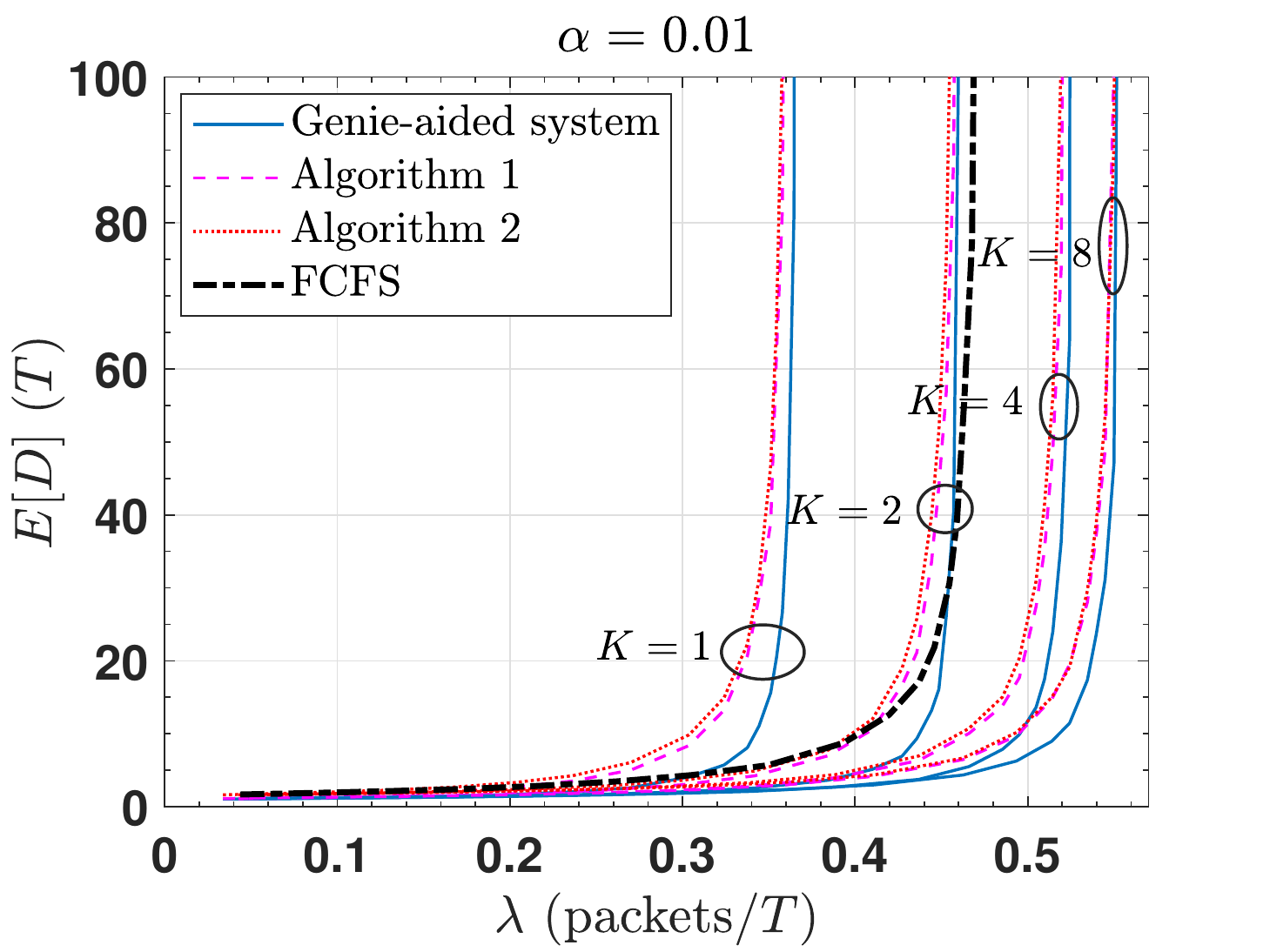}	\label{fig:fign7}}
		%		\subfigure[Throughput with large population sizes..]{
			%		\includegraphics[width=2.5in,height=1.9in]{fig1}	\label{fig:fign4}}
		%		\subfigure[The average idle period.]{	
			%		\includegraphics[width=2.5in,height=1.9in]{fig61}	\label{fig:fig7}}
		%\vspace{-0.3cm}
		%	\subfigure[Throughput of unslotted ALOHA system.]{	\includegraphics[width=2.5in,height=1.9in]{fig4}	
			%		\label{fig:fig3}}
		\caption{Performance of Bayesian backoff algorithm in unsaturated traffic scenario.}		\label{fig:fig5}
		%	\caption{Throughput with $T=1$ and various (small) population sizes.}
	\end{figure}
	
Table \ref{Tab2} summarizes the optimal $K$ and $\kappa$ for several $\alpha$'s, while in Figs. \ref{fig:fign16}-\ref{fig:fign17}, throughput is depicted as either $\alpha$, or $K$, or both increases. Using Lemma \ref{Lm:Lm1}, we obtain Table \ref{Tab2}: Given $\alpha$, we numerically find the best $K$ for the maximum throughput $\tilde{\tau}^*$ by trying all possible $K$.
 The observations from Table \ref{Tab2} can be summarized as follows: First, as $\alpha$ (an indicator of how fine micro-synchronism can be implemented in physical layer) increases, optimal $K$ (the number of TO's) of maximizing throughput becomes smaller; $\kappa$ (an indicator of how aggressively users are encouraged to access) decreases. Second, since the overall redundancy by $\alpha$ and $K$ is $\alpha(K-1)$, such a redundancy reduces  as $\alpha$ gets smaller. Third, as $\alpha$ decreases from $0.21$ to $0.01$, i.e., finer synchronism, the maximum throughput is improved from $5\%$ to $50\%$ with respect to S-ALOHA without TO's.

In Fig. \ref{fig:fign16}, for a large $K$, throughput decreases rapidly as $\alpha$ increases. Especially, for $\alpha\geq 0.28$, even though $K$ increases, we can not improve throughput better than S-ALOHA. It means that the duration of one TO should be less than 28\% of one packet transmission time; otherwise, inefficient. For a given $\alpha$, the optimal $K$ can be found in Fig. \ref{fig:fign17}. As $\alpha$ gets smaller, we can introduce more TOs for higher throughput. %Obviously, if a smaller $\alpha$ could be allowed in physical layer, a higher throughput can be expected.
%\end{figure*}

Returning to Figs. \ref{fig:fign3}-\ref{fig:fign6}, we can see throughput approximation by Corollary \ref{Cr:Cr1}, i.e., infinite population model, is also compared for $K=4$. In Fig. \ref{fig:fig5_new}, we further show the approximation in Corollary \ref{Cr:Cr1} for other $K$'s and $\alpha$'s against simulations and can observe good agreements between analysis and simulations. It can be seen that Poisson approximation made by Corollary \ref{Cr:Cr1} works reasonably well.

\begin{figure}[pt]\centering
%	\subfigure[Average number of users monitoring feedback per slot]{
	\includegraphics[width=3.3in,height=2.65in]{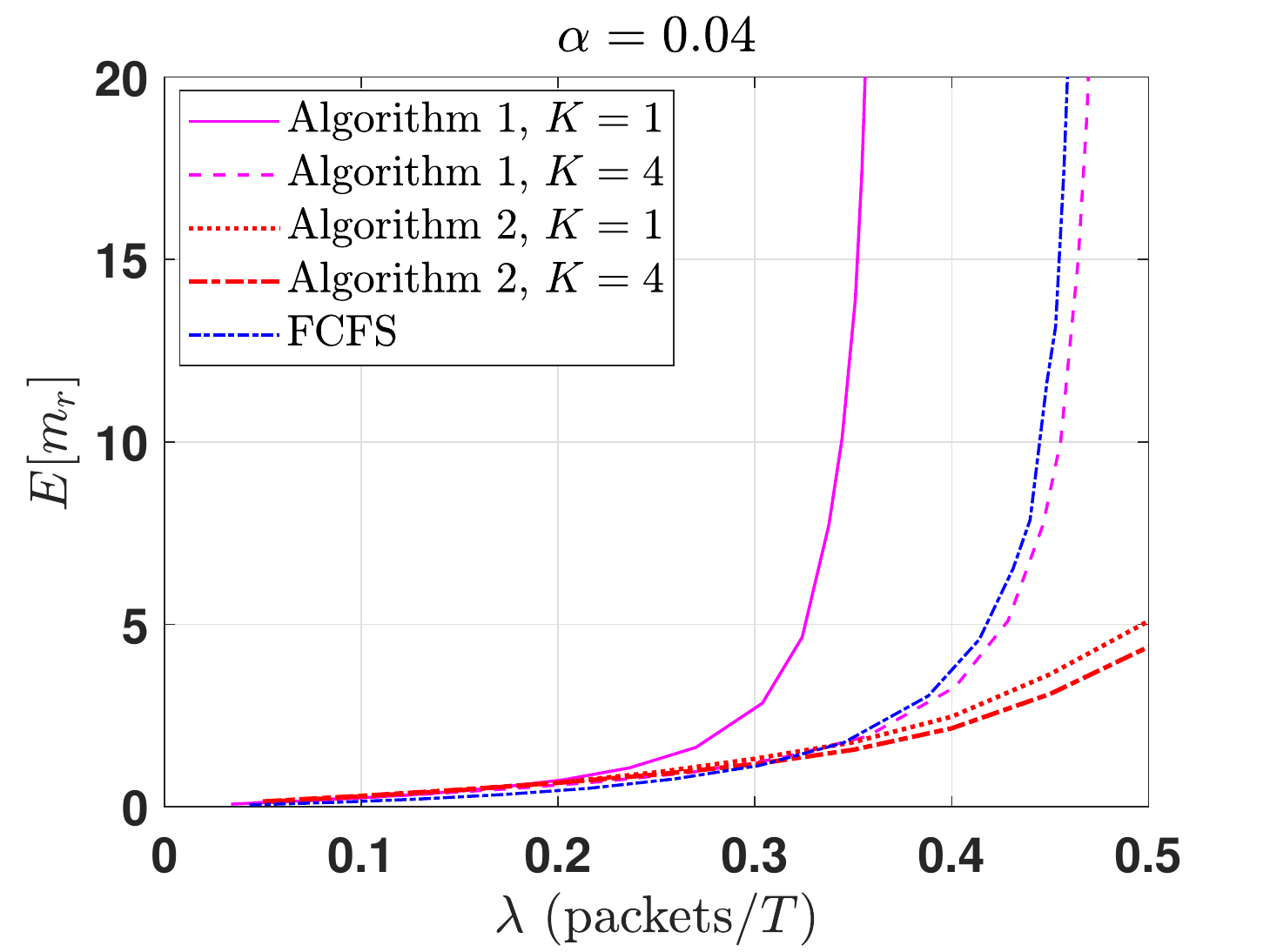}	
	\caption{Average number of users monitoring feedback per slot}\label{fig:fign14}
\end{figure}

%The proposed system with $\alpha=0.04$ and $K=4$ realizes the throughput corresponding to the best tree algorithm. 

%Fig. \ref{fig:fig20} shows the throughput-delay characteristic of unsatuated system. Good agreements between analysis and simulation show that Proposition \ref{Pr:Pr2} works well. As we have seen in the saturated system,  a larger $K$ increases throughput. 
 %\begin{figure}[pt]  \centering
%		\includegraphics[width=2.4in,height=1.84in]{throughput_delay}	
%	\caption{Throughput-delay characteristic of unsaturated S-ALOHA with TOs.}		\label{fig:fig20}
	%	\caption{Throughput with $T=1$ and various (small) population sizes.}
%\end{figure}
\begin{figure}[pt]  \centering
%\vspace{-1.5cm}
	\subfigure[Comparison of the a posteriori distributions.]{	\includegraphics[width=3.3in,height=2.65in]{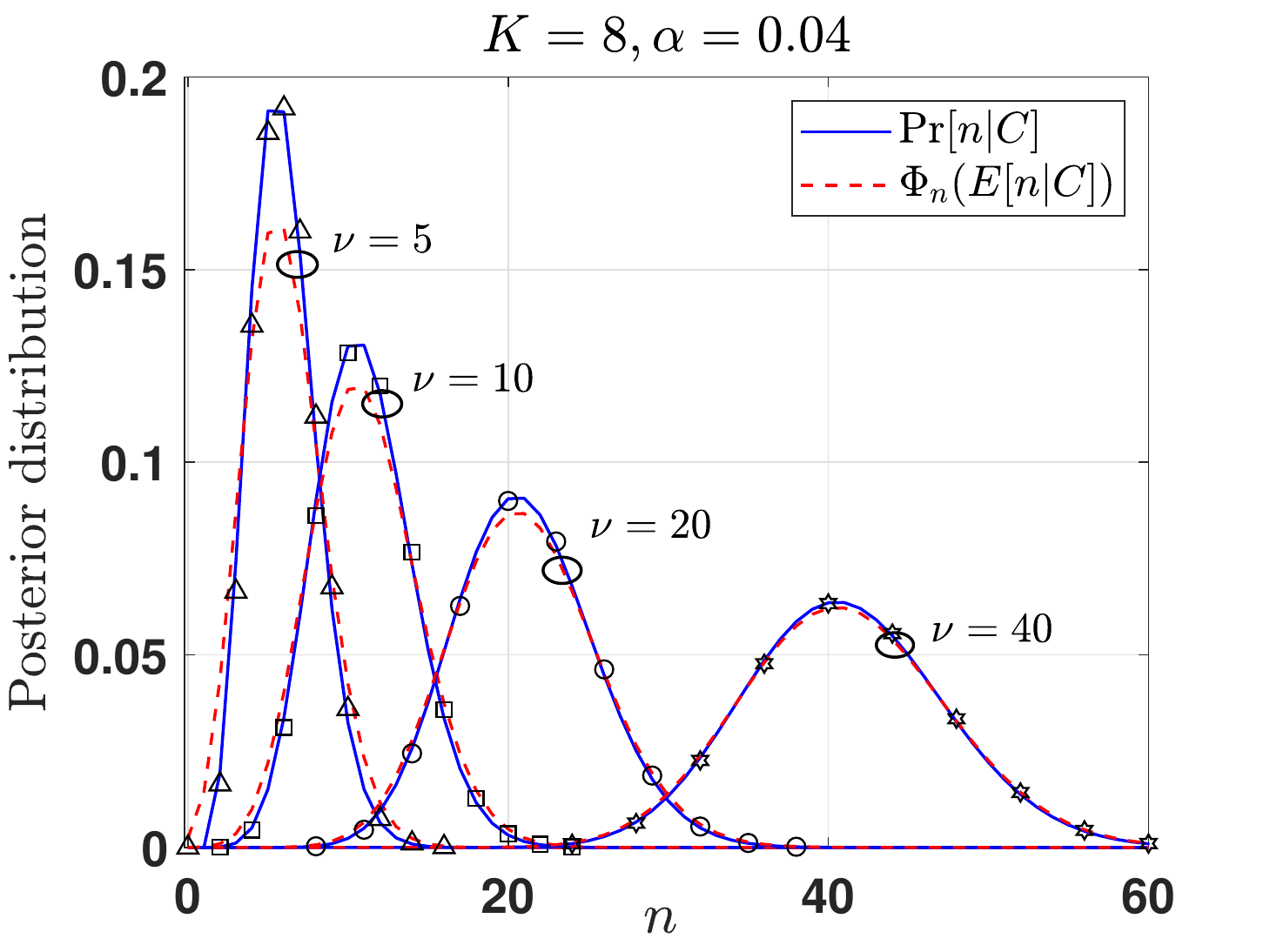}	\label{fig:fig7a}}\\
	\subfigure[Kull-back Leibler divergence.]{	\includegraphics[width=3.3in,height=2.6in]{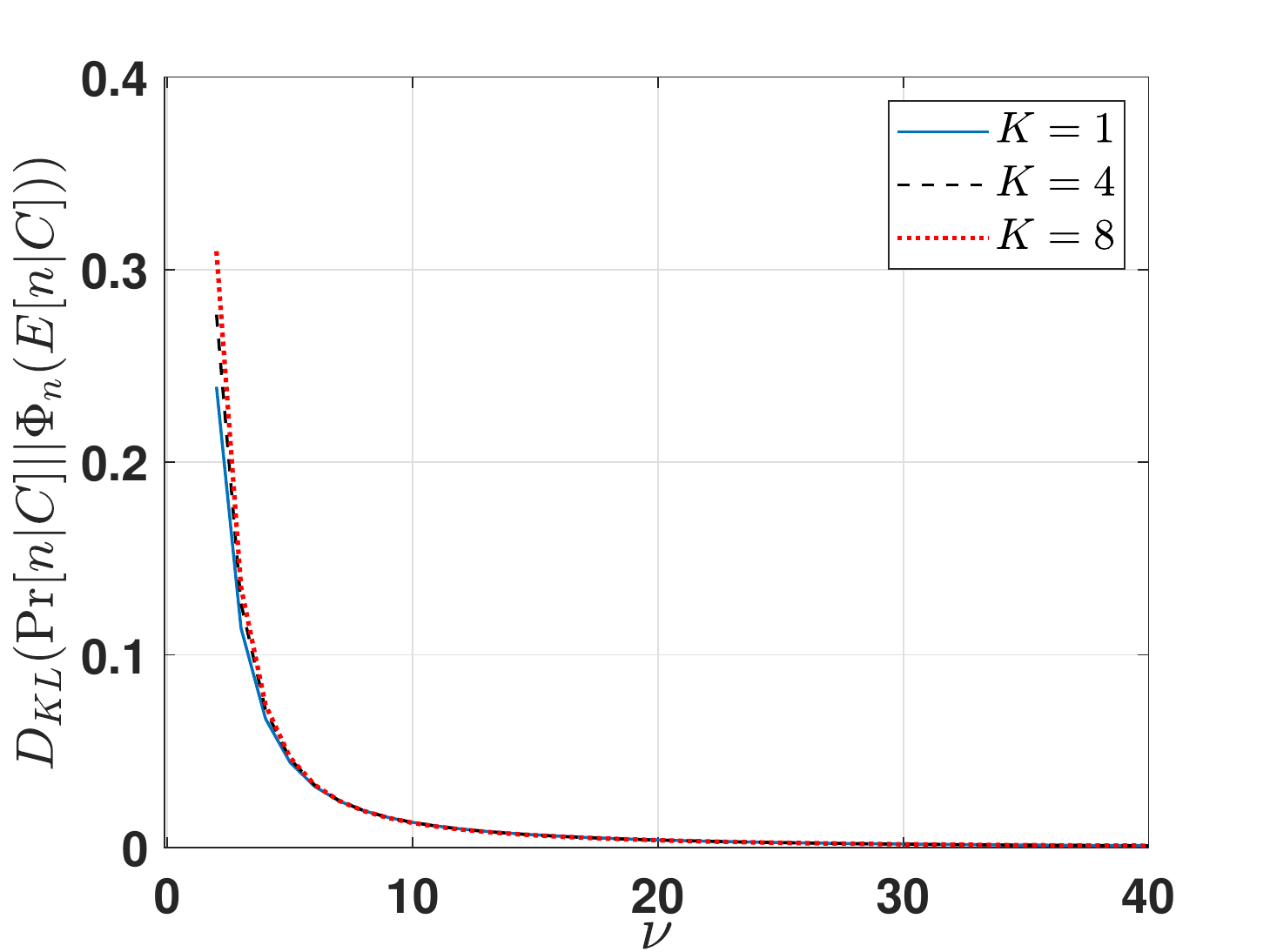}	\label{fig:fig7b}}\\
	\subfigure[Tracking the actual backlog size.]{	\includegraphics[width=3.3in,height=2.65in]{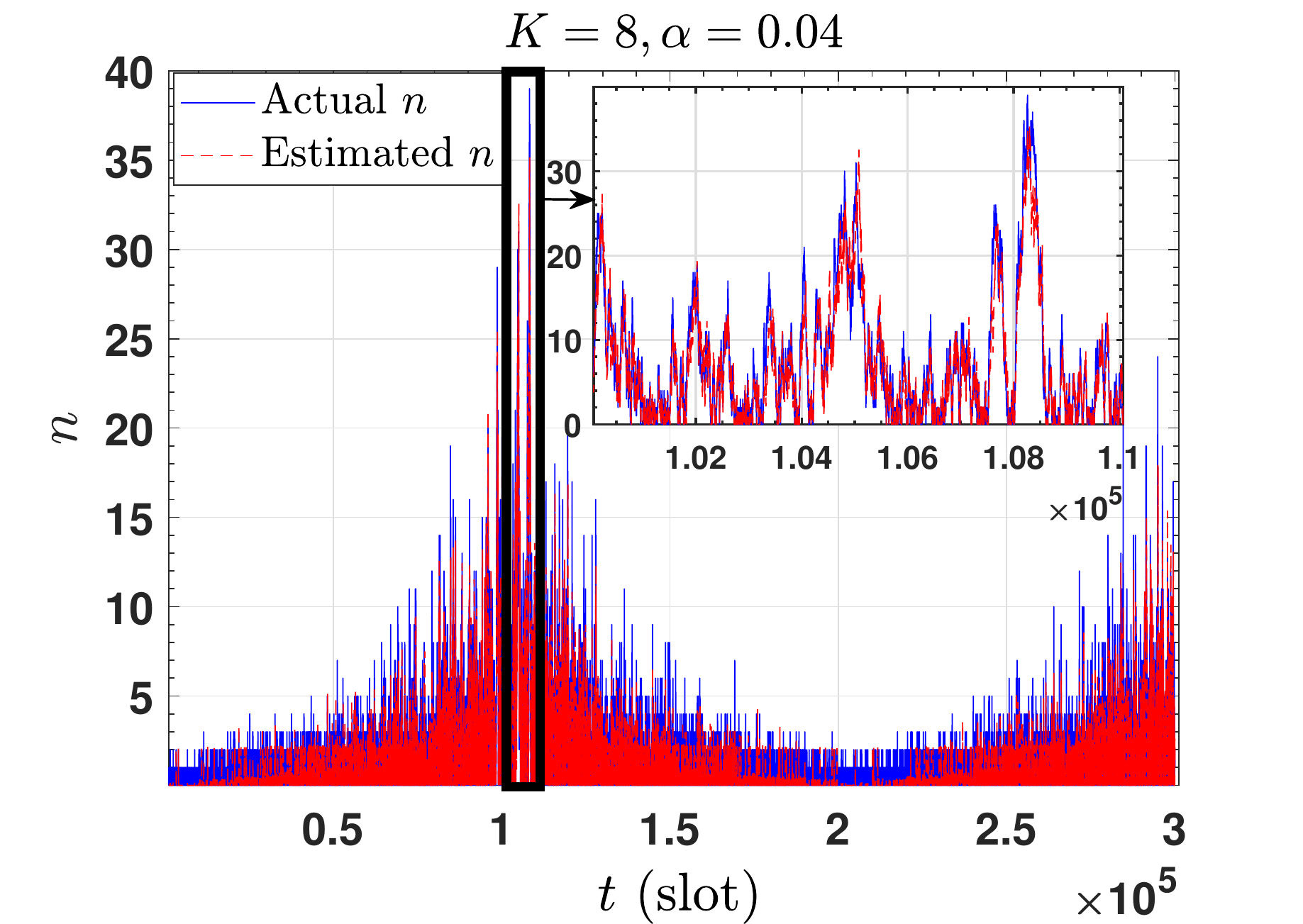}	\label{fig:fign9}}
	\caption{Behaviors of Bayesian algorithm.}		
\end{figure}

Figs. \ref{fig:fign7}-\ref{fig:fign8m} depict the average access delay of the Bayesian backoff algorithm, where $\lambda$ (packets/$T$) is the mean rate of new packet arrivals.  The lines in Fig.  \ref{fig:fig5} denote the simulation results. 
The proposed Bayesian backoff algorithm with retransmission probability $p_t^*$ and with uniform window $U_t$ are called algorithms 1 and 2, respectively.   For the proposed Bayesian bakcoff algorithm, we set the weighting factor $\theta = 0.99$. For comparison, we consider a genie-aided system \cite{Hajek}, where the BS knows the number of backlogged users at each slot perfectly. The difference is that the proposed Bayesian algorithm estimates the \emph{mean} of backlogged users and applies it to control $p$, whereas the genie-aided system makes use of  perfect knowledge in controlling $p$ to achieve the maximum throughput. Additionally, we compare FCFS splitting algorithm \cite{Bert}, whose maximum throughput is known as $0.487$ (packets/slot), i.e., the best tree algorithm known so far. It works as follows: At each slot, the BS announces an \emph{allocation interval}. The users whose packet \emph{arrived} (or generated) in the allocation interval can transmit at the next slot. Upon a collision, the previous allocation interval will be halved, whose first half becomes the next allocation interval. Upon consecutive collisions, the user whose packet arrived first in the initial allocation interval can be found by successive halving. If a slot is found idle during this procedure, from the fact that a collision occurred in the previous slot, the users belonging to the second half interval know certainly at least two users in the interval. They immediately split into the interval and the users in the first half are allowed to transmit. More details can be referred to \cite{Bert}. The weakness of FCFS algorithm for practical system is that the users' packet arrival time in the allocation interval should be infinitesimally divisible, i.e., no packet arrived at the same time. If the arrival time of two packets is identical, there is an indefinite loop of collisions. This is much more harder to implement physically than a small TO, i.e., $\alpha$. Moreover, during a collision resolution period, where successive interval halving occurs, the users should not lose any feedback information and other users with a new packet keep listening to the feedback every slot. Otherwise, the sequence of retransmissions governing FCFS algorithm will be broken, which deteriorates the overall performance. If the length of a TO goes very small, i.e., $\alpha\rightarrow 0$ as infinitesimally divisible allocation interval in FCFS algorithm,  the proposed algorithm can achieve $0.673$ throughput as shown in Corollary \ref{Cr:Cr1}, which is much better than $0.487$ of FCFS algorithm. In FCFS algorithm, if one successful transmission of the earliest arrival occurs, the next success of the next earliest arrival mostly follows during a contention resolution period. This is in fact an event of remarking the end of the ongoing contention resolution interval. In the proposed algorithm, the earliest and last arrivals in TOs can make a successful transmission and the maximum contention resolution period can be three slots.
\begin{figure}[pt]  \centering
%\vspace{-1.5cm}
	\subfigure[Number of active devices over time ($1000$ devices).]{	
		\includegraphics[width=3.3in,height=2.65in]{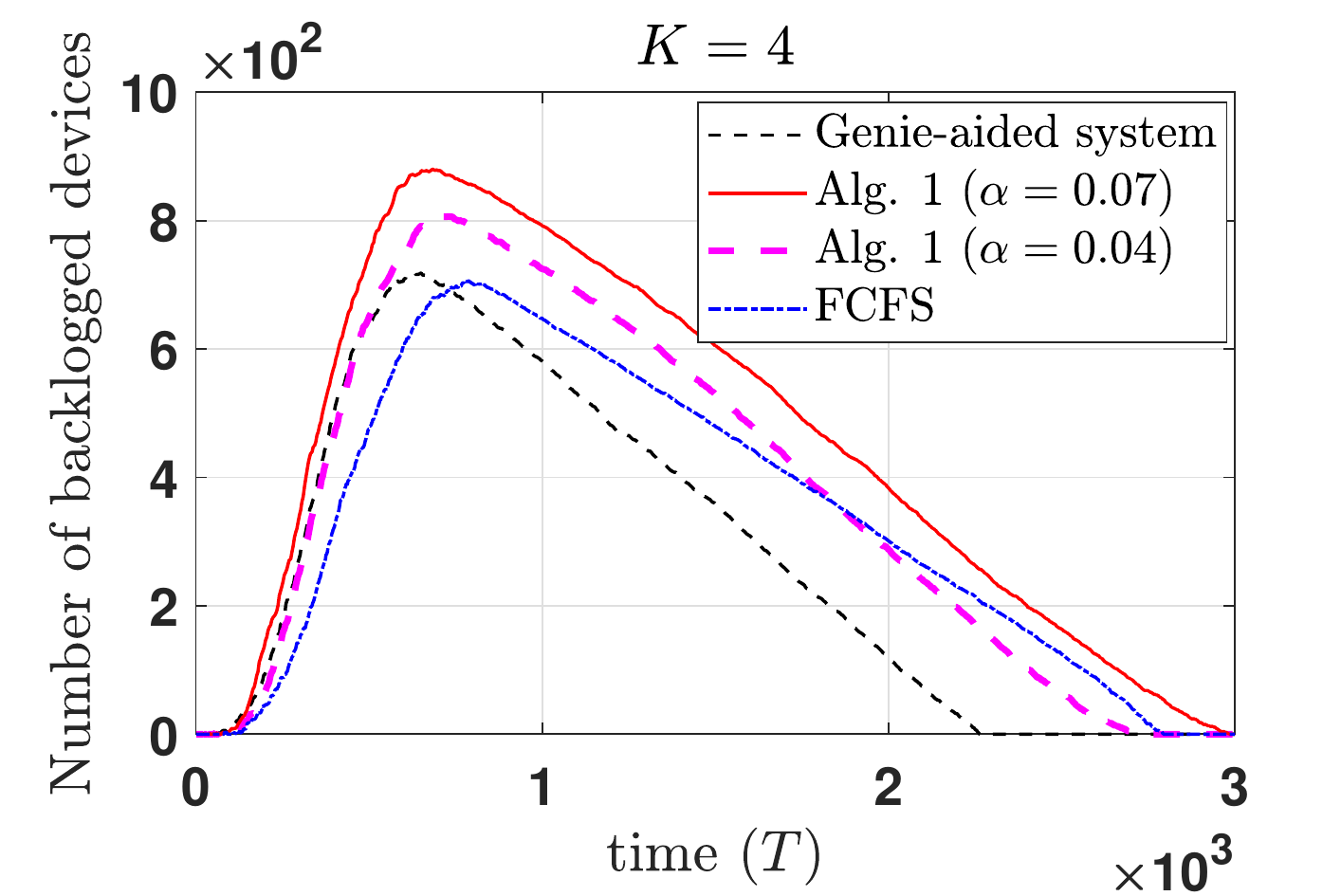}	\label{fig:fign10}}
	\subfigure[Number of active devices over time ($5000$ devices).]{	
		\includegraphics[width=3.3in,height=2.65in]{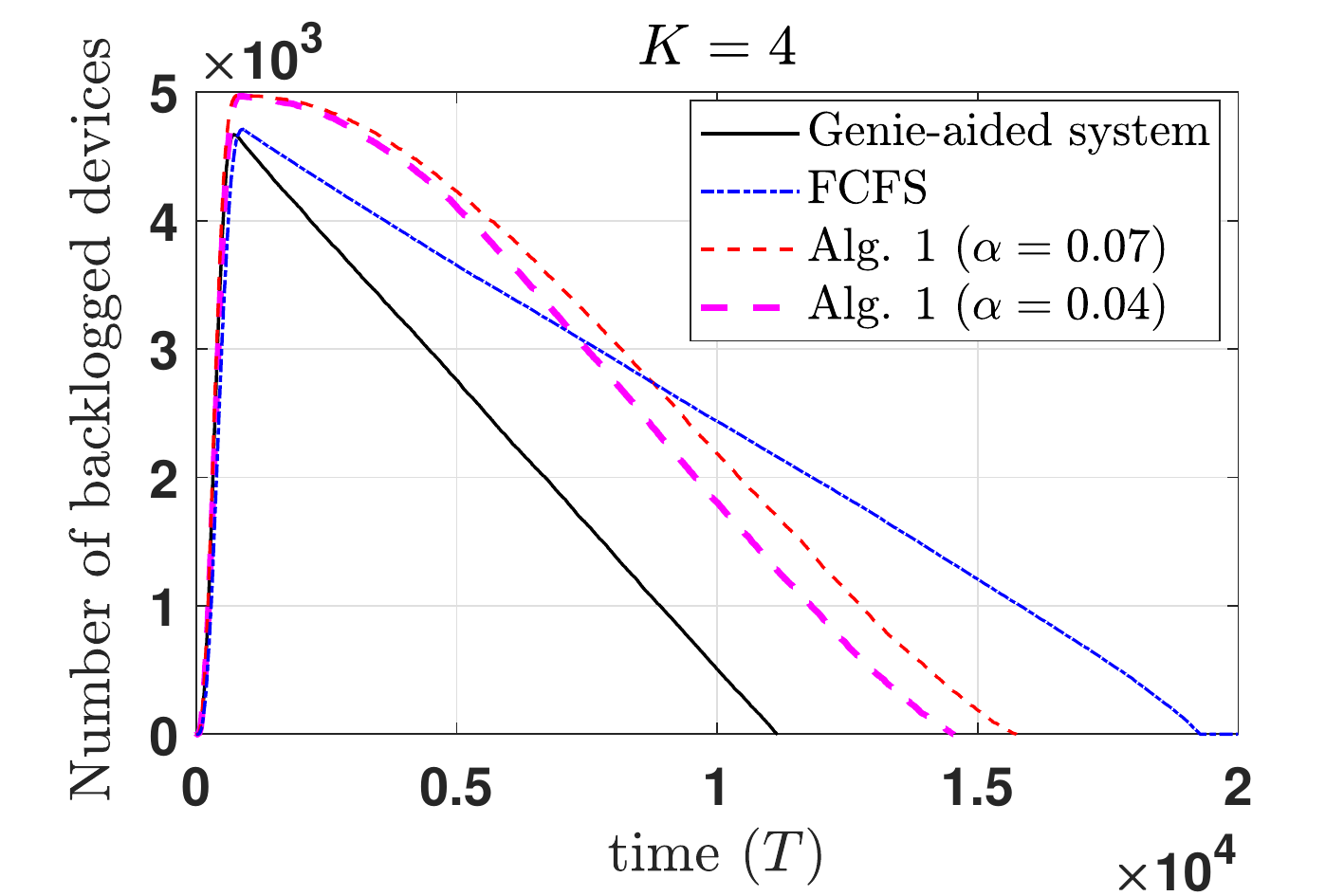}	\label{fig:fign11}}
	\subfigure[Total service time vs. the number of devices.]{	
		\includegraphics[width=3.3in,height=2.65in]{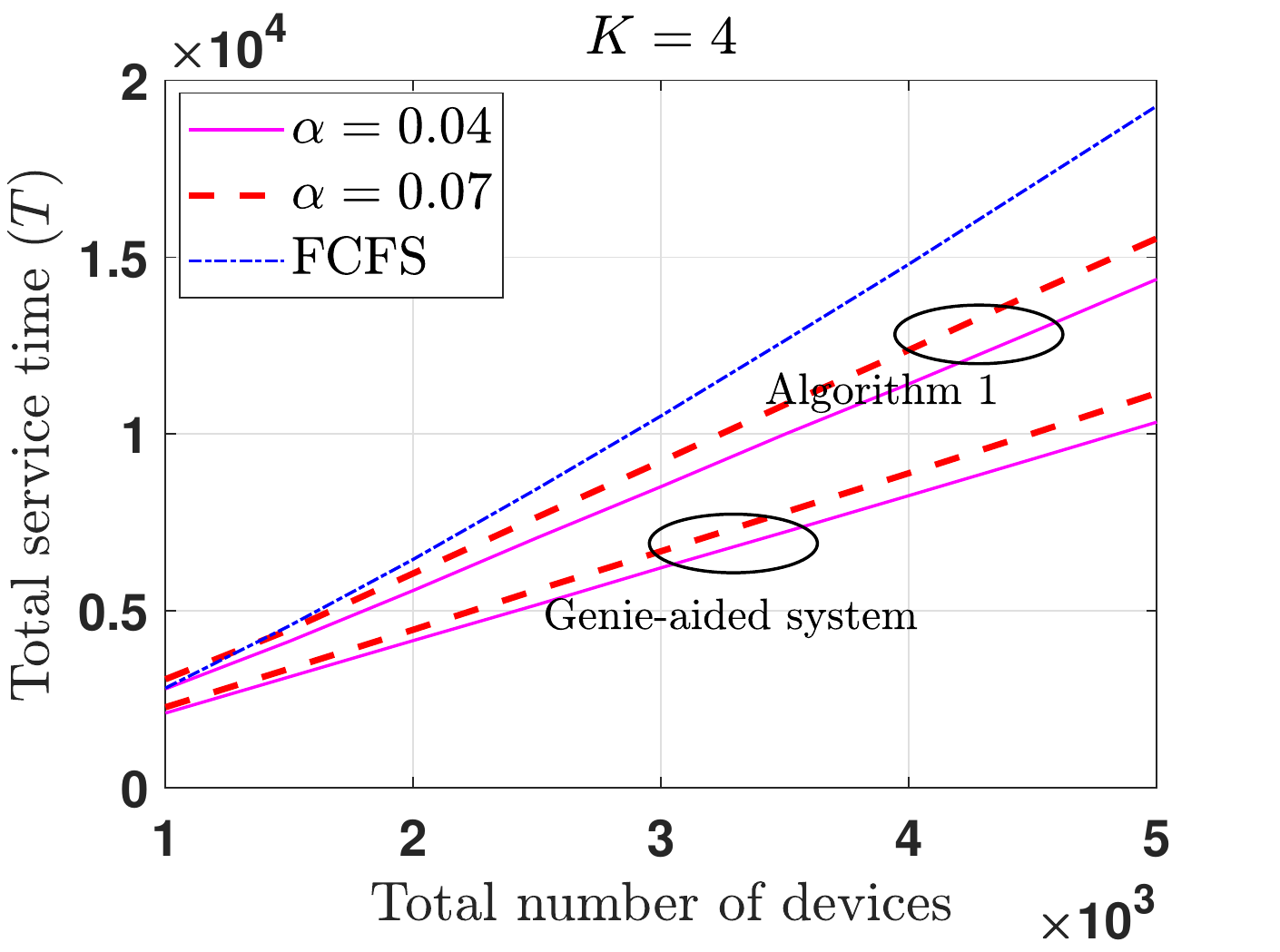}	\label{fig:fign12}}
	\caption{Performance with Beta-distributed traffic.}		\label{fig:fig9}
\end{figure}
In Fig. \ref{fig:fig5}, we can make the following observations: Referring to the maximum throughput presented in  Figs. \ref{fig:fign1} and \ref{fig:fign4}, as $\lambda$ gets close to $\tilde{\tau}^*$, it can be seen that the average access delay starts to grow without a bound. As mentioned before, notice that the system with  $K=1$ is S-ALOHA. Remarkably, algorithms 1 and 2 show almost identical performance in all the settings. For $\alpha=0.07$, the proposed algorithm is much better than S-ALOHA without a TO,  whereas FCFS algorithm seems better than the proposed algorithm. However, let us recall that it is not possible to implement FCFS algorithm in practice. As $\alpha$ becomes smaller from 0.07 in Fig. \ref{fig:fign8m} to 0.01 in Fig. \ref{fig:fign7}, the performance of the proposed Bayesian algorithm seems more sensitive to $K$ and becomes better than FCFS algorithm.

To show the advantage of Algorithm 2 over Algorithm 1 in terms of energy consumption for monitoring BS' broadcast message, we  show the average number of users monitoring the feedback from BS per slot, denoted by $\mathbb{E}[m_r]$, when $\alpha = 0.04$  in Fig. \ref{fig:fign14}. The more users monitor the feedback, the higher the energy consumption. It can be observed that Algorithm 2 reduces significantly the number of feedback-monitoring users, which is less than five.
With FCFS splitting algorithm, the number of the users monitoring the feedback is larger than in S-ALOHA with Algorithm 1 and four TOs. This is because all the users with FCFS algorithm should monitor it to find the end of each CRP.

In Figs. \ref{fig:fig7a} and \ref{fig:fig7b}, we check whether our Poisson assumption on \eqref{eq:eq41}, i.e., the a prior distribution, can be validated; recall that instead of \eqref{eq:eq41}, Poisson distribution with mean \eqref{eq:eq_collision} is used for deriving the algorithm. In Fig. \ref{fig:fig7a}, we compare the a posteriori distribution of \eqref{eq:eq41}, Poisson distribution and the empirical distribution from simulations. As $\nu$ increases, three distributions match well. In Fig. \ref{fig:fig7b}, we examine Kullback-Leibler divergence; it provides the information on how close together two distributions can be. If $\nu$ is more than six, it can be seen that two distributions seem quite close. Finally, Fig. \ref{fig:fign9} presents how well the Bayesian backoff algorithm can track the true backlog size well even for time-varying $\lambda$.   Instead of a constant $\lambda$, we vary $\lambda$ over time as follows: At the beginning, it is set to $\lambda = 0.039$ (packets/$T$).  Every $10^4$ slots,
$\lambda$ is raised by 0.039.  When $\lambda$ reaches 0.429, it is reduced by 0.039 again every $10^4$
slots.

In contrast with Poisson arrivals, we use the following traffic model to test traffic of massive IoT devices \cite{Jin}: For the system with a total of $N$ IoT devices, the devices are activated at time $x\in(0,T_A)$, whose probability density function (PDF) $f_B(x)$ is Beta distribution with parameters $a$ and $b$ as 
\begin{align}\label{eq:eqbeta}
	f_B(x) = \frac{x^{a-1}(T_A-x)^{b-1}}{T_A^{a+b-1}B(a,b)},
\end{align}
where $B(a,b)=\int_0^1x^{a-1}(1-x)^{b-1}dx$ is the Beta function. If $T_A$ consists of $I_A$ slots, the expected number of newly activated devices in the $i$-th slot is given by $\lambda_i=N\int_{t_{i-1}}^{t_i}F_B(x)dx$ for $i=1,2,\dots,I_A$. We use $a=3$, $b=4$ and $T_A=1000 T$, whereas the number of IoT devices $N$ is 1000 in Fig. \ref{fig:fign10} and 5000 in Fig. \ref{fig:fign11}, respectively. It is notable that the parameters of the proposed algorithm, e.g., $\kappa$, (Alg. 1 in Fig. \ref{fig:fig9}) do not change even with a different traffic model, because the throughput-optimal retransmission probability depends on the estimated average number of backlogged IoT devices in each slot. Fig. \ref{fig:fig9} shows the number of backlogged devices over time under the traffic model in \eqref{eq:eqbeta}. When it becomes zero, all the backlogged devices are cleared up. In Fig. \ref{fig:fign12}, the total service time is presented. As the number of devices increase, FCFS algorithm becomes worse, while  the total service time of the proposed algorithm  moderately increases.

%%%%%%%%%%%%%%%%%%%%%%%%%%%%%%%%%%%%%%%%%%%%%%%%%%%%%%%%%%%%%%%%%%%%%%%%%%%%%%%%%%%%%%%%%%%
%From Figs. \ref{fig:fign16} are the Figs. \ref{fig:fign17}, we can find there are optimal $K$ for each $\alpha$,
%Fig. \ref{fig:fign18} depicts the optimal $K$ for different $\alpha$, and Fig. \ref{fig:fign19} depicts the maximum throughput using the optimal $\nu p$, i.e. $\kappa$, and optimal $K$ for different $\alpha$.
%\begin{figure}[pt]  \centering
%	\subfigure[]{	
%		\includegraphics[width=2.4in,height=1.84in]{optimal_K}	\label{fig:fign18}}
%	\subfigure[]{	
%		\includegraphics[width=2.4in,height=1.84in]{max_throughput_alpha}	\label{fig:fign19}}
%	\caption{}		\label{fig:fig10}
%\end{figure}

\begin{Remark}
In practice, the BS may not be able to distinguish type-1 and type-2 collisions perfectly due to imperfect energy detection techniques.  By introducing the misdetection probability of type-1 collision $q$,  we can discuss its effect on the performance of the proposed system.  Although we can not relate this probability $q$ to how practical physcial layer techniques often fail, our generic modeling is still valuable to capture some insight to understand the effect of misdetection.  With this new parameter $q$, the throughput (packets/$T$) in \eqref{eq:eq61} can be modified by 
\begin{equation}
\tau_{n}(p, q)=\gamma \cdot \frac{\mathcal{B}_{1}^{n}(p)+2\left(1-q\right) \sum_{i=2}^{n} \sum_{j=1}^{K-1} \frac{i \mathcal{B}_{i}^{n}(p)}{K^{i}}(K-j)^{i-1}}{3-2q-2\left(1-q\right)\left(\sum_{i=0}^{1} \mathcal{B}_{i}^{n}(p)+\sum_{i=2}^{n} \frac{\mathcal{B}_{i}^{n}(p)}{K^{i-1}}\right)}. 
\end{equation}
When $n\rightarrow \infty$, the above throughput can be approximated as 
\begin{equation}
\tau_{\infty}\left(q\right) \approx \gamma \cdot \frac{\eta e^{-\eta}+\left(1-q\right)\frac{2 \eta}{K}\left[\frac{e^{-\frac{\eta}{K}}-e^{-\eta}}{1-e^{-\frac{\eta}{K}}}-(K-1) e^{-\eta}\right]}{3-2q-2\left(1-q\right) e^{-\eta}\left(K\left(e^{\frac{\eta}{K}}-1\right)+1\right)},
\end{equation}
where $\eta = np$.  By numerically searching for the optimal $\eta$ that maximizes $\tau_{\infty}(q)$, we plot Fig. \ref{fig:throughput_q} when $\alpha = 0.04$. For $q=1$, the BS cannot distinguish types of collisions. Thus,  we cannot expect throughput improvement compared to conventional S-ALOHA due to time resource additionally used for TOs.  However, as $q$ decreases, i.e., the BS is able to distinguish types of collisions, the throughput improvement becomes more significant.  It can be also observed that the value of $K$ shall be chosen suitably. 
\begin{figure}[pt]\centering
	\includegraphics[width=3.3in,height=2.65in]{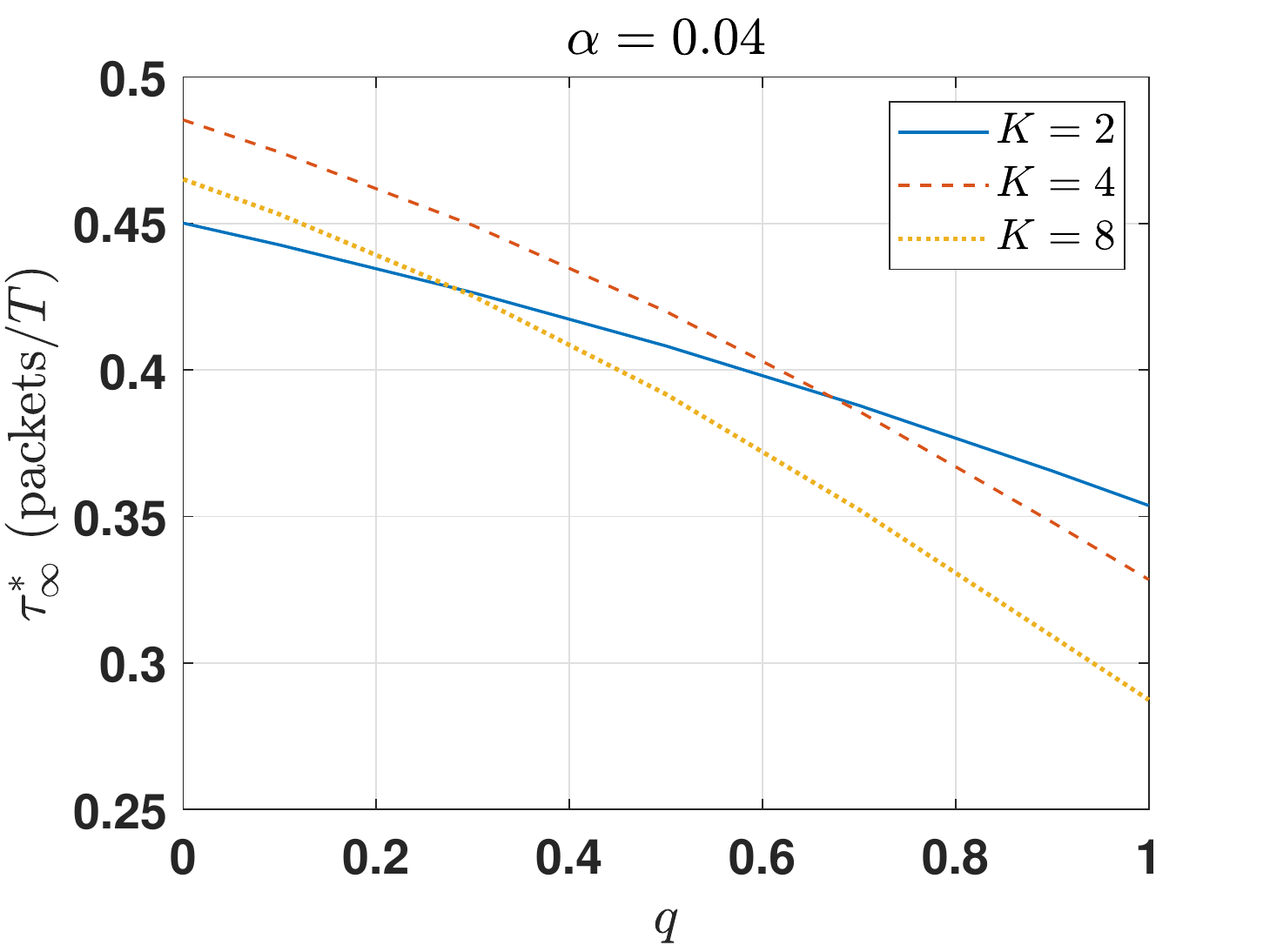}	
	\caption{Maximum throughput of $\tau_{\infty}(q)$ over $q$.}\label{fig:throughput_q}
\end{figure}

Fig. \ref{fig:delay_q} shows the average access delay of Algorithm \ref{algo1} for various $q$'s when  $\alpha = 0.04$ and $K=4$.  When running Algorithm \ref{algo1}, we used the optimal transmission probability by assuming $q=0$ because the BS may not know the value of $q$ a priori. We can observe that the proposed system still works properly, and the delay performance becomes better and the stable region for the arrival rate becomes larger when $q$ decreases.
\begin{figure}[pt]\centering
	\includegraphics[width=3.3in,height=2.65in]{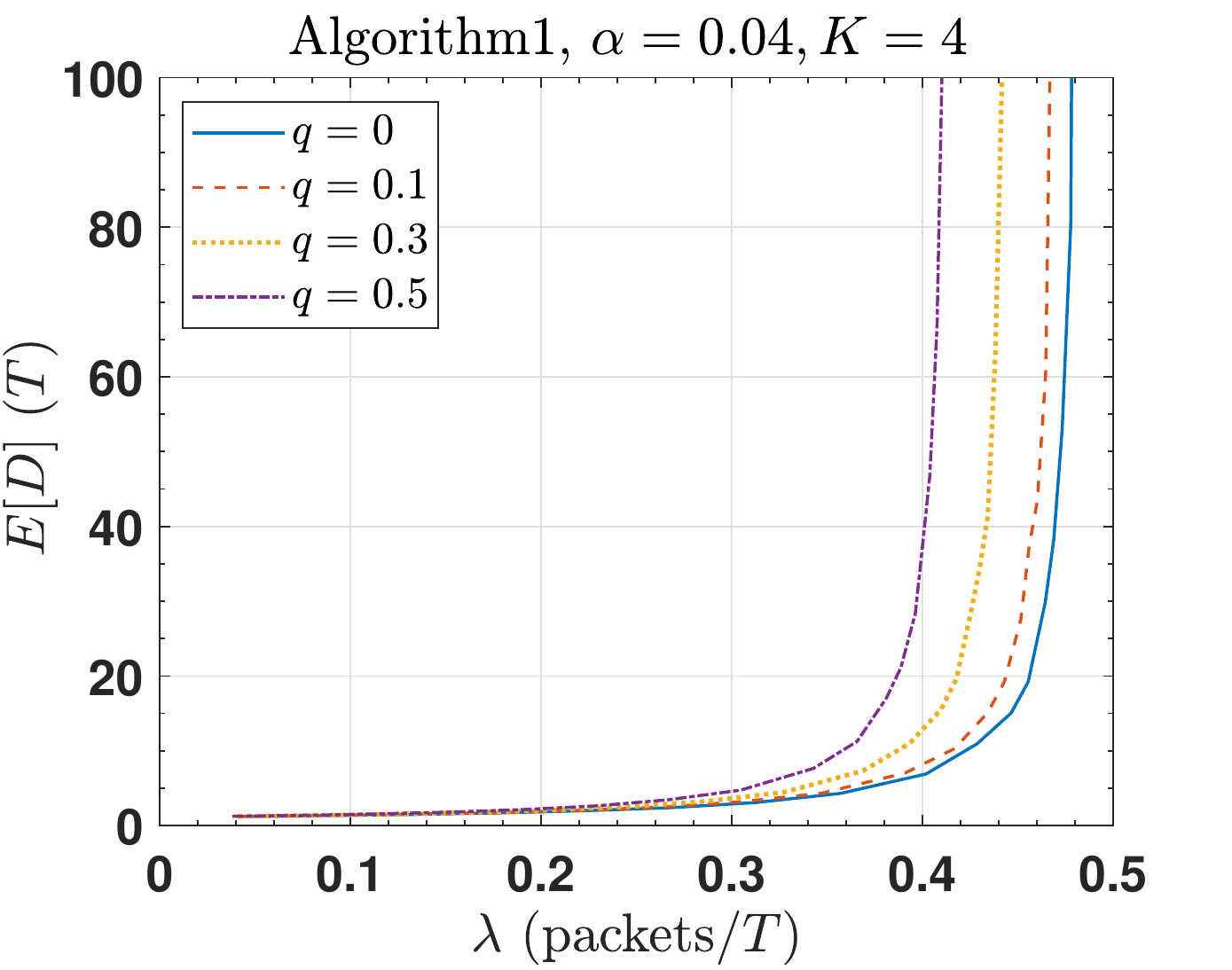}	
	\caption{Average access delay with different $q$.}\label{fig:delay_q}
\end{figure}
\end{Remark}

\section{Conclusion}\label{sec:con}
This work  proposed a novel S-ALOHA system with TOs for the next generation multiple access, where the TOs play a vital role in distinguishing the first and the last packets transmitted in a collision. We analyzed throughput and the average access delay under the saturated system so that throughput behavior and the maximum throughput with a fixed (re)transmission probability $p$ can be characterized according to the length and the number of TOs, i.e., $\alpha$ and $K$. Furthermore, a pair of the throughput-optimal system parameters  $(\alpha, K)$, which is found numerically, was presented when the estimated backlog size is assumed to have Poisson distribution.  Their effectiveness was confirmed by simulations with the traffic arrival patterns of Poisson and Beta distributions. Since the system runs in unsaturated condition in practice, it has been shown that the unsaturated system can be unstable with a fixed retransmission probability. To cope with this in practical versatile environments, we proposed a Bayesian backoff algorithm that controls (re)transmission probability or uniform window dynamically to realize the maximum throughput  under unsaturated traffic scenarios.  Results showed that S-ALOHA  system with TOs can achieve more than the known throughput barriers of S-ALOHA and tree algorithm, i.e., $0.3679$, and $0.487$. The throughput gain of the proposed system comes from a high precision of synchronization at TOs in physical layer. As future work, it is interesting to see how imperfect synchronization can harm the performance presented here.%, whereas the proposed Bayesian backoff algorithm. 

% or


\appendices
\section{Proof of Theorem \ref{Th:Th1}}\label{App1}
The throughput $\tau_n(p)$ is obtained based on renewal reward theorem in \cite{ross}. As in Fig. \ref{fig:figm2}, the channel goes through success, type-1 or 2 collision, and idle, randomly, which form a renewal cycle. Then, we can find the average reward obtained over the average length of the renewal cycle:
\begin{align}\label{eq:eq2}
	\tau_n=\frac{\mathbb{E}[R|n]}{\mathbb{E}[Z|n] / T},
\end{align}
where $\mathbb{E}[R|n]$  and ${\mathbb{E}[Z|n]}$ are the average reward and  the average renewal length given $n$ users.  The normalization by $T$ appears in the denominator because we set the unit of the throughput by packets per $T$ (sec). 

To find the average reward $\mathbb{E}[R|n]$, let us denote by $\Pr[S|n]$ the probability of success due to a single packet transmission in a slot, by  $\Pr[S_{i,c}|n]$ the probability of type-$i$ success under type-1 collision, and by $\Pr[\omega|n]$ the probability of a three-slot collision under type-1 collision. respectively. Then, $\mathbb{E}[R|n]$ is expressed as  
\begin{align}\label{eq:eq3}
	\mathbb{E}[R|n] = \Pr[ S |n ] +2\Pr[ S_{0,c}|n] + \sum_{i=1}^2\Pr[S_{i,c} | n]. 
\end{align}
Letting $\mathcal{B}_i^n(p) = {n\choose i}p^i(1-p)^{n-i}$, we get $\Pr[S|n]$ as 
\begin{align}\label{eq:eq4}
	\Pr[S|n]=\mathcal{B}_1^{n}(p).
\end{align}

To get other terms in \eqref{eq:eq3}, let us define $\Pr[S_{i} | n]$ for $i=1,2$ as
\begin{align}\label{eq:eq6}
	\Pr[S_{i}|n]=	\Pr[S_{0,c} |n] +\Pr[S_{i,c} |n].
\end{align}
Eq. \eqref{eq:eq6} accounts for the probability that the packet transmission in the $i$-th following slot under type-1 collision for $i\in\{1,2\}$ is successful. It does not matter whether the packet transmitted in the other slot (e.g., in the first slot if $i=2$) is success or not. Since using \eqref{eq:eq6} we have
\begin{align}\label{eq:eq6aa}
	2\Pr\left[ S_{0,c}|n\right] + \sum_{i=1}^2\Pr[S_{i,c}|n] = \sum_{i=1}^{2}\Pr[S_{i}|n],
\end{align}
we can rewrite \eqref{eq:eq3} as
\begin{align}\label{eq:eq6a}
	\mathbb{E}[R|n]=\Pr[S|n]+\sum_{i=1}^2\Pr[S_{i}|n].
\end{align}
In order to get $\text{Pr}\left[ S_{1}  |n \right]$, we first calculate the probability that the earliest packet transmission occurs at the $j$-th TO in type-1 collision while the others are transmitted after the $j$-th TO, given a total of $i$ packets transmitted in a slot. This is denoted by $\text{Pr}\left[ S_{1},j | i, n \right] $ and we get it as
\begin{align}\label{eq:eq8}
	\Pr[ S_{1} , j | i , n] ={\binom {i}{1}}\frac{1}{K}\left( 1-\frac{j}{K} \right) ^{i-1}.
\end{align}
In \eqref{eq:eq8}, it can be read that the number of ways of picking  up  one among $i$ users is ${i\choose 1}$, while the probability  that the chosen user  transmits to the $j$-th TO is $1/K$ due to random choice. Finally, we have the probability that the remaining $i-1$ users transmit to the remaining $K-j$ TOs after the $j$-th TO is $\left((K-j)/K\right)^{i-1}$. 

To get rid of $j$ in \eqref{eq:eq8}, let  us consider $\Pr\left[ S_{1}| i, n \right]$, i.e., the probability that the packet transmission in the first slot following after type-$1$ collision is successful, given that a total of $i$ packets have been transmitted in the previous collision slot. Summing this over $j$, we have
\begin{align}\label{eq:eq9}
	\Pr[ S_{1}| i , n] =\sum_{j=1}^{K-1}{\Pr}[  S_{1},j| i, n ] =\sum_{j=1}^{K-1}{\frac{i}{K^i}}\left( K-j \right) ^{i-1}.
\end{align}
Using \eqref{eq:eq9}, we can get $\Pr\left[S_{1} |n \right]$:
\begin{align}\label{eq:eq9m}
	\Pr\left[S_{1} |n  \right] =&\sum_{i=2}^n{\mathcal{B}_i^{n}(p)}\Pr\left[ S_{1}| i, n \right] \notag\\=&\sum_{i=2}^n{\sum_{j=1}^{K-1}{\frac{i\mathcal{B}_i^{n}(p)}{K^i}}}\left( K-j \right) ^{i-1}.
\end{align}
Similarly, to find the expression of $\Pr\left[ S_{2} |n  \right]$, let us consider  $\Pr\left[ S_{2},j| i, n \right] $, i.e., the probability that upon type-1 collision, given a total of $i$ packets transmitted in the slot, only one packet is  transmitted last at the $j$-th TO, while the other packets are transmitted before the $j$-th TO:
\begin{align}
	\Pr\left[ S_{2}, j| i, n \right] ={\binom {i}{1}}\frac{1}{K}\left( \frac{j-1}{K} \right) ^{i-1}.
\end{align}
As in \eqref{eq:eq8}, while we have ${i\choose 1}$ ways of choosing one of $i$ users randomly, this user transmits at the $j$-th TO with probability $1/K$. In addition,  the probability that the remaining $i-1$ users transmit to $j-1$ TOs before the $j$-th TO is $((j-1)/K )^{i-1}$. 

As in \eqref{eq:eq9m},  $\Pr\left[ S_{2} | i , n\right]$ denotes the probability that given a total of $i$ packets transmitted in the slot of type-1 collision, a successful packet transmission occurs in the  second following slot. This can be calculated as
\begin{equation}\label{eq:eq11}
	\begin{aligned}
		\Pr\left[ S_{2}| i , n\right]& =\sum_{j=2}^K\Pr\left[ S_{2},j | i, n \right]=\sum_{j=2}^K{\frac{i}{K^i}}\left( j-1 \right) ^{i-1}\\
		&=\sum_{j=1}^{K-1}{\frac{i}{K^i}}\left(K-j \right) ^{i-1}=\Pr\left[S_{1} | i, n \right].
	\end{aligned}
\end{equation}
Therefore, we further have $\Pr\left[ S_{2} | n\right] = \Pr\left[ S_{1} | n\right]$. 

So far we have obtained the average reward $\mathbb{E} [R|n ]$ in \eqref{eq:eq2}.  Let us find the average renewal length $\mathbb{E} [Z|n ]$ in \eqref{eq:eq2}. To do this, let $\Pr[I |n  ]$ be the probability that a slot is found idle given $n$ users in Fig. \ref{fig:figm2},   while $\Pr[C_i | n]$ indicates the probability of type-$i$ collision. In addition, when  $\mathbb{E}[Z| x, n]$ is the expected time period given event $x$, $\mathbb{E} [Z|n ]$ can be expressed as
\begin{equation}
	\begin{aligned}\label{eq:eq12}
		&	\mathbb{E} [Z|n ] = \mathbb{E}[Z| I, n]\Pr[ I| n] +\mathbb{E}[Z| S, n]\Pr[ S|n ] \\&~~~+\mathbb{E}[Z|  C_2, n] \Pr[ C_2 | n] +\mathbb{E}[Z| C_1, n] \Pr[ C_1| n].
	\end{aligned}
\end{equation}
From Fig. \ref{fig:figm2} it is not difficult to see that
\begin{align}
	\mathbb{E}[Z|I, n]=	\mathbb{E}[Z|S, n ]=	\mathbb{E}[Z|C_2, n]=T_s,
\end{align}
and since type-1 collision takes three slots, we have
\begin{align}
	\mathbb{E}[Z|C_1, n]=3T_s.
\end{align}
Therefore, $\mathbb{E}[Z|n]$ in \eqref{eq:eq12} can be expressed as
\begin{align} \label{eq:eq_renewal_length}
	\mathbb{E}[Z|n] = &T_s\big(\Pr[ I |n] +\Pr[ S|n ] +\Pr[ C_2|n] +3 \Pr[ C_1|n]  \big),
\end{align}
where $\Pr[S|n]$ is given in \eqref{eq:eq4}. 

Since an idle slot occurs when no one transmits, we have 
\begin{align} \label{eq:eq_idle}
	\Pr[I|n]=\mathcal{B}_0^{n}(p)	.
\end{align}
To get $\Pr[ C_2|n]$, let $\Pr[ C_2| i , n] $ denote the probability of type-2 collision, i.e., probability that all packets are transmitted in the same TO, given a total of $i$ packets transmitted in the slot:
\begin{align}
	\Pr[ C_2| i, n ] ={\binom {K}{1}}\left( \frac{1}{K} \right) ^i=\frac{1}{K^{i-1}}.
\end{align}
Then, we can get $\Pr\left[ C_2 | n\right]$:
\begin{align}\label{eq:eq18}
	\Pr\left[ C_2 | n \right] =\sum_{i=2}^n\frac{1}{K^{i-1}}\mathcal{B}_{i}^{n}(p).
\end{align}
Finally, we need to find $\Pr[C_1|n]$ in \eqref{eq:eq12}. Let $\Pr[\mathbb{C}|n]$ denote the probability of the overall collision probability, i.e., $\Pr[\mathbb{C}|n]=\Pr[C_1\cup C_2|n]= \Pr[C_1|n]+\Pr[C_2|n]$ and 
\begin{align}\label{eq:eq19}
	\Pr[\mathbb{C}|n] + \Pr[I|n]+ \Pr[S|n]=1.
\end{align}We then get
\begin{align} \label{eq:eq_C1}
	\Pr[C_1|n]=&1-\Pr[I|n]-\Pr[S|n]-\Pr[C_2|n]\notag\\=&1-\mathcal{B}_0^n(p)
	-\mathcal{B}_1^n(p)-\sum_{i=2}^n\frac{1}{K^{i-1}}\mathcal{B}_{i}^{n}(p).\end{align}
Based on \eqref{eq:eq4}, \eqref{eq:eq_idle},  \eqref{eq:eq18}, and \eqref{eq:eq_C1}, we can obtain $\mathbb{E}[Z|n]$ in \eqref{eq:eq_renewal_length}.  This completes the proof.	
\section{Proof of Corollary \ref{Cr:Cr1}}\label{App2}
Since a binomial distribution is approximated by a Poisson distribution as $n$ grows large, i.e., $\eta=np$ as $n\rightarrow \infty$ and $p\rightarrow 0$, let us start with replacing $\mathcal{B}_i^n(p)$ with $\Phi_i(\eta)$ when $n$ gets large:
\begin{equation}
	\begin{aligned}
		&\tau_{\infty} \approx \gamma\frac{\Phi_{1}(\eta)+2 \sum_{i=2}^{\infty} \sum_{j=1}^{K-1} \frac{i \Phi_{i}(\eta)}{K^{i}}(K-j)^{i-1}}{3-2\left(\sum_{i=0}^{1} {\Phi_{i}}(\eta)+\sum_{i=2}^{\infty} \frac{\Phi_{i}(\eta)}{K^{i-1}}\right)}\\
		&=\gamma\frac{\Phi_{1}(\eta)+2 \sum_{j=1}^{K-1} \frac{\eta }{K} e^{-\frac{\eta j}{K}} \sum_{i=2}^{\infty} {\Phi_{i-1}}\left(\eta\left(1-\frac{j}{K}\right)\right)}{3-2\left(\sum_{i=0}^{1} {\Phi_{i}}(\eta )+ K e^{\frac{(1-K)\eta}{K} }\sum_{i=2}^{\infty} {\Phi_{i}}\left(\frac{\eta}{K}\right)\right)}\\
		&=\gamma\frac{\Phi_{1}(\eta)+2 \sum_{j=1}^{K-1} \frac{\eta}{K} e^{-\frac{\eta j}{K}}\left[1-\Phi_{0}\left(\eta \left(1-\frac{j}{K}\right)\right)\right]}{3-2\left(\sum_{i=0}^{1} {\Phi_{i}}(\eta)+K e^{\frac{(1-K)\eta}{K}}\left(1-\sum_{i=0}^1\Phi_{i}\left(\frac{\eta}{K}\right)\right)\right)} 
		\\
		&=\gamma\frac{\eta e^{-\eta}+\frac{2 \eta}{K}\left[\sum_{j=1}^{K-1} \left(e^{-\frac{\eta}{K}}\right)^j-(K-1) e^{-\eta }\right]}{3-2 e^{-\eta }\left(K \left(e^{\frac{\eta}{K}}-1\right)+1\right) },
	\end{aligned}
\end{equation}
which can be further simplified as \eqref{eq:eq22}. 

\section{Proof of Proposition \ref{Pr:Pr1}}\label{App3}
To find the upper-bound, let us suppose a system with $K=\infty$, and $\alpha=0$; that is, $(K-1)\alpha=0$, and $\gamma=1$. Since this system has infinite number of TOs, the probability that users choose the same TO is 0, whereas the length of each TO is zero. For $n$ saturated users, so we have $\text{Pr}[C_2|n]=0$, and $\text{Pr}[S_{0,c}|n]=\text{Pr}\left[ C_1|n \right]=\text{Pr}\left[ \mathbb{C}|n \right]$. Thus, throughput is expressed as
\begin{equation}
	\begin{aligned}
		\tau^o_n &=\frac{\Pr\left[ S|n \right] +2\Pr\left[ \mathbb{C} |n\right]}{T_s\left( \Pr\left[ I|n \right] +\Pr\left[ S|n \right] +3\Pr\left[ \mathbb{C} |n\right] \right)/T }\\
		&=\frac{2-2\mathcal{B}_{0}^{n}(p)-\mathcal{B}_{1}^{n}(p)}{3-2\left(\mathcal{B}_{0}^{n}(p)+\mathcal{B}_{1}^{n}(p)\right) }.
	\end{aligned}
\end{equation}
As $n\rightarrow \infty$,  we have
\begin{equation}
	\begin{aligned}
		\tau^o_\infty =\frac{2-2\Phi_{0}(np)-\Phi_{1}(np)}{3-2[\Phi_{0}(np)+\Phi_{1}(np)] },%\\
		%			&=\frac{2-(2+np)e^{-np}}{3-2e^{-np}(1+np) }
	\end{aligned}
\end{equation}
which completes the proof. 
\section{Proof of Lemma \ref{Lm:lm2}}\label{App4}
	This can be examined by Foster-Lyapunov theorem \cite{Bert}. Let us define the following variables: Let $X_t$ be the number of backlogged users at the beginning of  an open slot $t$. In addition, $B_t$ and $A_{L_t}$ denote the number of backlogged users who make successful transmissions, and the number of new packet arrivals during $L_t$ slots. Then, the evolution of $X_t$ can be expressed as 
\begin{align}\label{eq:eq40a}
	X_{t+L_t} = X_t - B_t + A_{L_t},
\end{align}
where $L_t$ can be either one or three, depending on the channel outcome. In \cite{Bert}, the system is said to be stable  (or the Markov process of $X_t$ is ergodic) if
\begin{align} \label{eq:eq_drift}
	&\mathbb{E}[X_{t+L_t}- X_t | X_t = m] <0.
\end{align}
Plugging  \eqref{eq:eq40a} into \eqref{eq:eq_drift}, we get
\begin{align} \label{eq:eq_drifta}
	&\mathbb{E}[X_{t+L_t}- X_t | X_t = m] = \mathbb{E}[-B_t + A_{L_t} | X_t = m] \notag \\
	&~~~~= - \mathbb{E}[B_t|X_t = m] + \mathbb{E}[A_{L_t} | X_t = m] < 0.
\end{align}
In fact, $\mathbb{E}[B_t|X_t = m]$ is the average number of successful transmissions given $m$ backlogged users:
\begin{align} \label{eq:eq_Bt}
	\mathbb{E}[B_t | X_t = m] = \mathbb{E}[R |m].
\end{align}
Moreover, $\mathbb{E}[A_{L_t}|X_t = m]$ is the mean packet arrivals to the system: 
\begin{align}\label{eq:eq_At}
	&\mathbb{E}[A_{L_t}|X_t = m] =\lambda \mathbb{E}[Z|m] / T.
\end{align}
where $\lambda$ (packets/$T$) is the average packet arrival rate to the system.  Plugging \eqref{eq:eq_Bt}  and \eqref{eq:eq_At} into  \eqref{eq:eq_drift}, we have 
\begin{align}\label{eq:eq53}
	\lambda &< \frac{\mathbb{E}[R |m]}{\mathbb{E}[Z |m] / T }= \tau_m(p).
\end{align}	
If we use $p_m^*$ (throughput-optimal retransmission probability for $m$ backlogged users) for $p$, the right-hand side (RHS) of \eqref{eq:eq53} is the maximum throughput given $m$. If we find its minimum, \eqref{eq:eq_drift} always holds. 

\section{Proof of Corollary \ref{Cr:Cr4}} \label{App5}
Suppose that there can be a state, denoted by $m^\star$, from which we have $\lambda>\tau_m(p)$ for $m>m^\star$. Since all the states of  Markov process $X_{t+L_t}$ in Lemma \ref{Lm:lm2} can communicate, $X_{t+L_t}$ can visit that state at least once.  After that, $X_{t+L_t}$ keeps moving higher states. Then, there is no steady-state probability distribution of $X_{t+L_t}$. We need to show that as $m\rightarrow \infty$ with fixed $p$, $\tau_{\infty}$  in \eqref{eq:eq22}  goes to zero due to $\eta=mp\rightarrow \infty$:
\begin{align}
	\tau_{\infty}&\approx \gamma \cdot \frac{\eta e^{-\eta }+\frac{2 \eta}{K}\Bigg[\frac{e^{-\frac{\eta}{K}}-e^{-\eta}}{1-e^{-\frac{\eta}{K}}}-(K-1) e^{-\eta}\Bigg]}{3-2 e^{-\eta }\left(K \left(e^{\frac{\eta}{K}}-1\right)+1\right)}\notag\\
	&=\gamma \cdot \frac{\eta e^{-\eta }+\frac{2}{K}\Bigg[\frac{\eta e^{-\frac{\eta}{K}}-\eta e^{-\eta}}{1-e^{-\frac{\eta}{K}}}-(K-1) \eta e^{-\eta}\Bigg]}{3-2 \left(K \left(e^{-\frac{\eta(K-1)}{K}}-e^{-\eta }\right)+e^{-\eta }\right)}\notag\\
	%&= \gamma \cdot \frac{0+\frac{2}{K}\Big[\frac{0-0}{1-0}-(K-1) 0\Big]}{3-2 \left(K \left(0-0\right)+0\right)}\notag\\
	&=\gamma \cdot \frac{0}{3}=0,
\end{align}
where L'Hopital's rule has been used and $e^{-\eta}\rightarrow 0$ for $\eta\rightarrow\infty$.
%% use section* for acknowledgment
%\ifCLASSOPTIONcompsoc
%  % The Computer Society usually uses the plural form
%\else
%  % regular IEEE prefers the singular form
%\fi

\ifCLASSOPTIONcaptionsoff
  \newpage
\fi

\vfill

% Can be used to pull up biographies so that the bottom of the last one is flush with the other column.

\end{document}